\documentclass[]{aastex701}
\usepackage{CJK}
\usepackage{orcidlink}
\usepackage{verbatim}
\usepackage{enumerate}
\usepackage{graphicx}
\usepackage{gensymb}
\usepackage{multirow}
\usepackage{makecell}
\usepackage{ulem}
\usepackage{CJK}
\usepackage{natbib}
\usepackage{lineno}
\linenumbers
\usepackage{amsmath,amsfonts,amssymb,mathrsfs,graphicx,color}
\usepackage{xcolor}
\usepackage{color, colortbl}
\usepackage{slashed}
\usepackage[utf8]{inputenc}
\usepackage{float}
\usepackage{soul}

\newcommand{\eg}{{\it e.g.}}
\newcommand{\eq}{Eq.~}
\newcommand{\Fig}{Fig.~}
\newcommand{\Sec}{Sec.~}
\newcommand{\Tab}{Tab.~}

\newcommand{\equ}[1]{\eq(\ref{equ:#1})}
\newcommand{\figu}[1]{\Fig\ref{fig:#1}}
\newcommand{\tableref}[1]{\Tab\ref{tab:#1}}
\newcommand{\secref}[1]{\Sec\ref{sec:#1}}

\newcommand{\prince}{{\sc PriNCe}}
\newcommand{\xmax}[0]{X_\text{max}}

\newcommand{\MeanXmax}{$\langle X_\mathrm{max}\rangle$}
\newcommand{\SigmaXmax}{$\sigma(X_\mathrm{max})$}

\newcommand{\uh}{UHECR}
\newcommand{\td}{TDE}
\newcommand{\tds}{TDEs}

\newcommand{\n}{neutrino}

\newcommand{\msun}{M_{\odot}}

\begin{document}
\begin{CJK*}{UTF8}{gbsn}

\title{Ultra-High-Energy Cosmic Rays from Neutrino-Emitting Tidal Disruption Events}

\correspondingauthor{Pavlo Plotko}
\email{pavlo.plotko@desy.de, https://plotkopavlo.com}


\author[0000-0001-6975-5186]{Pavlo Plotko}\affil{Deutsches Elektronen-Synchrotron DESY, 
Platanenallee 6, 15738 Zeuthen, Germany}\affil{Deutsches Zentrum f\"{u}r Astrophysik/German Center for Astrophysics (DZA) , Postplatz 1, 02826 G\"{o}rlitz}
\email{pavlo.plotko@desy.de}

\author[0000-0001-7062-0289]{Walter Winter}\affil{Deutsches Elektronen-Synchrotron DESY, 
Platanenallee 6, 15738 Zeuthen, Germany}
\email{walter.winter@desy.de}

\author[0000-0002-9253-1663]{Cecilia Lunardini}\affil{Department of Physics, Arizona State University, 450 E. Tyler Mall, Tempe, AZ 85287-1504 USA}
\email{cecilia.lunardini@asu.edu}

\author[0000-0003-0327-6136]{Chengchao Yuan (袁成超)}\affil{Deutsches Elektronen-Synchrotron DESY, 
Platanenallee 6, 15738 Zeuthen, Germany}
\email{chengchao.yuan@desy.de}

\shorttitle{UHECRs from TDEs}
\shortauthors{Plotko, Winter, Lunardini, Yuan}

\begin{abstract}
We revisit the Ultra-High-Energy Cosmic Ray (UHECRs) production in Tidal Disruption Events (TDEs) in the light of recent neutrino-TDE associations. We use an isotropically emitting source-propagation model, which has been developed to describe the neutrino production in AT2019dsg, AT2019fdr, and AT2019aalc. These TDEs have strong dust echoes in the infrared range, which are potentially linked with the neutrino production. A mechanism where neutrinos originate from cosmic ray scattering on infrared photons implies cosmic rays in the ultra-high energy range, thus suggesting a natural connection with the observed UHECR. 
We extrapolate the three TDE associations to a population of neutrino- and UHECR-emitting TDEs, and postulate that these TDEs power the UHECRs. We then infer the source composition, population parameters, and local rates that are needed to describe UHECR data. We find that UHECR data point towards a mix of light to mid-heavy injection isotopes, which could be found, e.g., in oxygen-neon-magnesium white dwarfs, and to a contribution of at least two groups of TDEs with different characteristics, dominated by AT2019aalc-type events. The required local TDE rates of  ${\mathcal O}(10^2)~\mathrm{Gpc^{-3} \, yr^{-1}}$, however, are more indicative of the disruption of main sequence stars. We propose an enhanced efficiency in the acceleration of heavier nuclei that could address this discrepancy. The predicted diffuse neutrino fluxes suggest a population of astrophysical neutrino sources that can be observed by future radio neutrino detection experiments. The derived source parameters are consistent with those expected from the individual neutrino observations. 
\end{abstract}

\section{Introduction}\label{sec:intro}
 
Tidal Disruption Events (TDEs) are astrophysical transients originating from  stars that are tidally destroyed by a 
supermassive black hole (SMBH). The accretion of bound stellar materials could power a month to year-long flare \citep{1988Natur.333..523R,1989IAUS..136..543P}, whose spectrum extends from radio to X-ray frequencies. There have currently been ~170 TDEs, TDE candidates and TDE suspects \citep{2011ApJ...741...73V, Hammerstein_2022, masterson2024newpopulationmidinfraredselectedtidal}. They have been identified by the Zwicky Transient Facility \citep[ZTF,][]{2019PASP..131a8002B}, the  Wide-field Infrared Survey Explorer \citep[{WISE},][]{2010AJ....140.1868W}, and X-ray/radio surveys, such as eROSITA \citep{2021MNRAS.508.3820S} and Very Large Array Sky Survey \citep[VLASS,][]{2020PASP..132c5001L}. Remarkably, the correlation study of an $\sim$200 TeV IceCube neutrino event IC-191001A and optical transients has revealed that a TDE, AT2019dsg, could be its most likely astrophysical counterpart \citep{2021NatAs...5..510S}. The subsequent joint analyses of IceCube neutrino events and electromagnetic signals have discovered more TDE candidates with potential neutrino associations, including AT2019fdr \citep{2022PhRvL.128v1101R}, AT2019aalc \citep{2021arXiv211109391V}, two dust-obscured TDE candidates \citep{2023ApJ...953L..12J}, and AT2021lwx \citep[located at a relatively high redshift $z\simeq1.0$,][]{2024arXiv240109320Y}.

The potential correlations between TDEs and astrophysical neutrinos provide the smoking-gun evidence of cosmic ray accelerations since high-energy neutrinos are generated from the interactions between energetic nuclei and ambient radiation/matter backgrounds. However, for TDEs, the sites for particle accelerations and interactions are still unclear. Many models including relativistic jets \citep{2011PhRvD..84h1301W,2016PhRvD..93h3005W,2017MNRAS.469.1354D,2017PhRvD..95l3001L,Senno:2016bso}, accretion disks \citep{2019ApJ...886..114H}, wide-angle outflows/hidden winds \citep{2020ApJ...904....4F}, and tidal stream interactions \citep{2015ApJ...812L..39D,2019ApJ...886..114H} have been proposed as the origin of the non-thermal electromagnetic (EM) and neutrino emission from TDEs, which could potentially explain these TDE-neutrino coincidences \citep[e.g.,][]{2020PhRvD.102h3028L,2020ApJ...902..108M, 2021NatAs...5..472W, 2022MNRAS.514.4406W,  2021NatAs...5..436H,2023ApJ...948...42W,2023ApJ...956...30Y}. 

The non-detection of $\gamma$-rays from these neutrino-coincident TDEs by the \emph{Fermi} Large Space Telescope (LAT) exerts stringent constraints on the TDE source parameters. In particular, the recent investigation of electromagnetic radiation induced by secondary particles of hadronic (e.g., $pp$ and $p\gamma$) processes demonstrates that the production of Ultra-High-Energy Cosmic Rays (UHECRs) with energies $\gtrsim10^{18}-10^{19}$ eV is permitted in an isotropic radiation region of radius $\gtrsim10^{16}-10^{17}$ cm without exceeding the $\gamma$-ray upper limits \citep{2023ApJ...956...30Y}.  Moreover, it was observed that the neutrino-TDE associations are accompanied by strong dust echoes in the infrared wavelength band, and that the arrival times of the coincident neutrinos were close to the peak of the infrared luminosity (which is delayed with respect to the OUV blackbody peak). This led to proposing cosmic ray scattering on infrared photons as a primary neutrino production mechanism, where cosmic-ray primary energies in the UHECR range are needed to exceed the photopion production threshold \citep[e.g.,][]{2023ApJ...948...42W,2023ApJ...956...30Y,2024arXiv240109320Y}. These conclusions support the idea that TDEs can be promising UHECR sources, and thus offer a possible answer to the long-standing problem of the origin of the observed UHECR flux   \citep{2009ApJ...693..329F,2014arXiv1411.0704F,Zhang:2017hom,2018A&A...616A.179G,2018NatSR...810828B,2023JCAP...11..049P}. 

The origin of UHECRs is another puzzle in astroparticle physics \citep{Alves_Batista_2019}. It is challenging to associate the measured CRs with a specific source class,  \citep{Capel_2019, Abdul_Halim_2024, CRPHYS_2014__15_4_339_0, Shaw_2022}. Even with theoretical modeling of promising source classes, it remains unclear which specific source class is responsible for UHECRs, including  Gamma-Ray Bursts  \citep[GRBs,][]{Heinze_2020, Boncioli_2019, 10.1111/j.1365-2966.2011.19590.x, Murase_2006, Samuelsson_2019, Vietri_1995, PhysRevLett.75.386} , Active Galactic Nuclei \citep[AGNs, ][]{Rodrigues_2018, 10.1093/mnras/stx498, Dermer_2010}, Starburst galaxies \citep{refId0, PhysRevD.97.063010, PhysRevD.60.103001}, and Galaxy clusters \citep{10.1093/mnras/286.2.257, Murase_2008, Kotera_2009}. Sources that produce UHECRs also might emit neutrinos, which contribute to the high-energy diffuse neutrino background. Therefore, the joint consideration of UHECR data and the IceCube diffuse neutrino background would provide valuable insights into the physical environments (e.g., the abundance of intermediate and heavy elements), the acceleration mechanisms (e.g., how the acceleration efficiency depends on the nuclei mass), and redshift distributions of UHECR sources.

In this study, we revisit the potential role of TDEs as sources of UHECRs, motivated by the recent identification of a possible link between three observed TDEs/candidates (AT2019dsg, AT2019fdr, and AT2019aalc) and high-energy neutrino detections by IceCube. While AT2019dsg is well established as a TDE, the classification of AT2019fdr and AT2019aalc remains uncertain, since both occurred in active galaxies where AGN activity can mimick tidal disruption flares; hence, these events are often more broadly referred to  as ``accretion flares'' \citep{van_Velzen_2021}. In addition, the interpretation of AT2019fdr as a superluminous supernova could also account for the coincident neutrino detection \citep{Pitik:2021dyf}. Based on the models presented in \citet{2023ApJ...948...42W}, we aim to explore the range of parameters that describe TDEs by analyzing various factors, including the type of disrupted stars, the size of the production region, and the maximum energies of CRs. By examining these parameters, we aim to identify scenarios that reproduce the UHECR data and satisfy bounds from the diffuse neutrino flux observations. To achieve this, we develop detailed computational models of cosmic ray emission in TDEs using the NeuCosmA \citep{Baerwald_2011,H_mmer_2010} software and then model the transport of UHECRs from sources to Earth using the \prince\ code \citep{Heinze:2019}. By comparing the results with the Akaike Information Criterion, we determine which scenarios best fit current observational data, leading to an improved understanding of TDEs as potential sources of UHECRs.

The paper is structured as follows: In \secref{model}, we develop a model of TDEs using AT2019dsg, AT2019fdr, and AT2019aalc as templates. We identify a set of parameters that will be used in a fit to the data, and determine the ranges of these parameters that are allowed by the observations of these TDEs and TDE candidates. In \secref{TDE_population}, we review the current understanding of TDEs as a population of sources, and introduce our scenarios of reference of the UHECR-neutrino-emitting population of TDEs. Using a fitting procedure, we compare these scenarios with the 
data on the UHECR spectrum and composition from Auger, and present a detailed analysis of the best-fit scenario and how well it compares to the data. In \secref{resulst_enhancement} we elaborate on the question of the consistency with the observed UHECR composition, and discuss a possible enhancement factor that would favor the acceleration of heavier elements in TDEs. In \secref{discussion}, we discuss the broader consistency of our findings with theoretical and observational expectations, identify areas of tension, and outline directions of future study. Throughout the paper, we use CGS units unless otherwise specified.

\section{Modeling of individual TDE\lowercase{s}}
\label{sec:model}
In this section, we develop and apply a detailed numerical model to simulate the production of UHECRs and neutrinos from TDEs. 

\subsection{TDE source model}
\label{subsec:source_model}

The observed neutrino-TDE associations share certain common properties: relatively high blackbody luminosities in the optical-UV range, strong dust echoes in the infrared range --  which are delayed by $\Delta T \sim {\mathcal O}(10^2)$ days with respect to the blackbody peaks  --  and the neutrinos arriving with comparable delays as well. X-rays were also observed in all these cases. The dust echoes are typically interpreted as re-processed radiation from higher energies, where the delays $\Delta T$ are indicative of the size of the dust torus surrounding the source. 
Considering that the neutrino's time delays are comparable with $\Delta T$, and the suggestion that neutrino-emitting TDEs and TDEs with strong dust echoes might be the same population \citep{2021arXiv211109391V}, it was proposed that the IR photons might be the main target for the neutrino production \citep[e.g.,][]{2023ApJ...948...42W,2023ApJ...956...30Y} - although a contribution from other wavelengths must be present. This requires that the cosmic ray primary nuclei of mass number $A$ and charge number $Z$ have high enough energies $E_A$ to exceed the photo-pion production threshold.  The (kinematical) photopion threshold is given by $E_N > m_\pi \, (2 \, m_p + m_\pi) c^4/(4 \, \varepsilon)$, where $E_N$ is the nucleon energy, $\varepsilon$ is the target photon energy, $m_p$ is the proton mass, $m_\pi$ is the pion mass, $c$ is the speed of light; here heads-on collisions are assumed at threshold. 

In a superposition ansatz, where the energy per nucleon $E_N \simeq E_A/A$, the threshold condition can be translated into one for the rigidity $R \equiv  p_A \, c/(Z \, e) \simeq E_A/(Z \, e)$, where $p_A$ is the nuclear momentum and $e$ is the elementary charge:
\begin{equation}
R \gtrsim R_{\mathrm{th}} \, \quad \text{where} \quad R_{\mathrm{th}} \simeq  7 \, 10^8 \, \mathrm{GV}\left(\frac{\varepsilon }{0.1 \, \mathrm{eV}} \right)^{-1} \, \frac{A}{Z} \, .
\label{equ:rmax}
\end{equation}
Thus, for heavy isotopes, where $A \sim 2 \, Z$ in the stability valley, one needs $R \gtrsim  10^9 \, \mathrm{GV}$ for target photons in the infrared range, which is independent of the composition to a first approximation. Consequently, the maximal rigidity (which we will use instead of the maximal energy)  has to exceed this value, which means that the primary cosmic rays will be in the  UHECR energy range.
Since the neutrino energy $E_\nu \simeq 0.05 \, E_N \simeq 0.05 \, E_A/A \simeq 0.05 \, R \, e \,  Z/A$, their interaction will produce neutrinos in excess of about 35~PeV. The emergent discrepancy between the high neutrino energies from UHECR interactions and the observed (mean) neutrino energies below a PeV is a challenge applicable to many source classes for which an UHECR connection is postulated; an exception are sources with strong (kG) magnetic fields such as Gamma-Ray Bursts, where synchrotron cooling of the secondary muons, pions and kaons can break the direct connection between UHECR and neutrino energies, see e.g. \citet{2012arXiv1201.5462W}; we will return to this issue in \Sec~\ref{sec:discussion}.

The emergent toy model picture is then a production region of size $r$ given by the size of the dust torus, within which the infrared photons produce a quasi-isotropic background photon field. The UHECRs will be accelerated somewhere within this region --  such as winds interacting with each other or with the circum-TDE material, off-axis or choked jets -- or even in some more compact regions (even though the model probably does not capture all contributions to the neutrino production then). 

 We adopt the quasi-isotropic emission model in \citet{2023ApJ...948...42W} (model M-IR)
 which uses the measured luminosities and temperatures as closely as possible. The interesting photon targets are the blackbody optical-UV emission (OUV, temperature $k_BT_{\mathrm{OUV}}\sim \mathrm{eV}$ with $k_B$ being the Boltzmann constant), the time-delayed, modelled dust echo in the infrared range (IR, temperature $k_BT_{\mathrm{IR}} \sim 0.1 \, \mathrm{eV}$), and X-rays probably from the innermost regions (temperature $k_BT_{\mathrm{X}} \sim 100 \, \mathrm{eV}$). These photon spectra are modeled by black (gray) body spectra with the individual TDE parameters and time evolutions, see \figu{evolution} (black dashed-dotted curves) for two examples. From the (observed) time delays $\Delta T$ of the IR versus OUV peaks we obtain the size of the dust torus $r_{\rm IR} \sim c \, \Delta T/2 \simeq 10^{17} - 10^{18} \, \mathrm{cm}$; within this radius quasi-isotropic IR photons are expected. Depending on the size of the region $r$ (inferred from $r_{\rm IR}$), on the magnetic field $B$, and on the cosmic-ray energy $E_A$, this volume can be calorimetric to CRs --  i.e., magnetically confined CRs lose their energy by radiation losses --  or even optically thick to them.  

We assume that cosmic-ray nuclei  $^Z$A  are accelerated in an acceleration zone which we do not model explicitly, but instead (similar to AGN blazar models) parameterize it by its injection luminosity $L_A$ and maximal rigidity $R_{\mathrm{max}}$, where $E_{A,\mathrm{max}} = Z \, eR_{\mathrm{max}}$. The cosmic-ray injection spectrum $J_A$ (density differential in energy, time and volume) is assumed as
\begin{equation}
 J_A = J_0 \, f_{A} E_A^{-2} \, \exp \left( - \frac{E_A}{E_{A,\mathrm{max}}} \right) \, ,    \label{equ:inj}
\end{equation} 
where $f_A$ is defined as the fraction of the energy density injected as element $A$ above $10^9 \, \mathrm{GeV}$ and $J_0$ is a normalization factor that will be derived from \equ{LA} below.
The non-thermal injection luminosity $L_A$ of nuclei with mass number $A$ is assumed to follow the mass accretion rate $\dot M$ as  
\begin{equation}
L_A = \varepsilon_{\mathrm{diss}} \, \dot M \, c^2 = \frac{4}{3} \, \pi \, r^3 \int\limits_{E_{A,\mathrm{min}}}^{\infty} E \, J_A \, dE ,
\label{equ:LA}
\end{equation}
where we use $\varepsilon_{\mathrm{diss}} \simeq 0.2$ as fiducial parameter;\footnote{As a consequence, $\varepsilon_{\mathrm{diss}}$ assumes the role of the ``baryonic loading'', a parameter commonly used in GRB or AGN models. Note that this formula assumes the injection of a single nuclear species ($f_A = 1$); later we will use mixed injection compositions whose total luminosities adds up to $\varepsilon_{\mathrm{diss}} \, \dot M \, c^2$.} we will later discuss its degeneracy with other parameters in \Sec~\ref{sec:discussion}. 
The mass accretion rate is assumed to have a peak value of $100 \, L_{\mathrm{Edd}}/c^2$ \footnote{The mass accretion rate $100L_{\rm Edd}/c^2$ is obtained from order-of-magnitude estimations \citep[e.g.,][]{2023ApJ...956...30Y}. Noting that during the TDE peak accretion phase, the accretion rate could be super-Eddington, e.g., $\dot M/\dot M_{\rm Edd}=\zeta\sim10-100$ \citep{2018ApJ...859L..20D}, where $\dot M_{\rm Edd}$ is the Eddington accretion rate that can be related to the Eddington luminosity $L_{\rm Edd}$ via $\dot M_{\rm Edd}=L_{\rm Edd}/(\eta_{\rm rad}c^2)$ using the radiation efficiency $\eta_{\rm rad}\sim0.1-0.01$. Therefore, we estimate $\dot M=(\zeta/\eta_{\rm rad})L_{\rm Edd}/c^2\sim\mathcal{O}(100)L_{\rm Edd}/c^2$, and adopt the fiducial value $100L_{\rm Edd}/c^2$ for consistency with previous studies.} and to trace the observed OUV luminosity time evolution.
The minimal energy $E_{A,\mathrm{min}} = \gamma_\mathrm{min} \, m_p \, c^2 \, A \propto Z$ with $\gamma_\mathrm{min} \simeq 1$, which means that all non-thermal nuclei are picked up (injected) at the same minimal gamma factor (or minimal rigidity) -- which is relevant if the mass fraction of the progenitor is to be translated into the non-thermal contribution. Note that the maximal ridigity can be directly converted into the maximal injection energy for a single injection isotope. Later, we will, however, combine different injection isotopes under the assumption of equal $R_{\mathrm{max}}$, which is the simplest possible assumption. This assumption is obtained if the maximal energy is limited by the size of the accelerator (confinement condition from balancing Lorentz and centrifugal forces). If other processes (such as synchrotron losses or photodisintegration) limit the maximal energy, they will change this relationship, and it may not apply anymore.

In summary, our key parameters, which we vary, are: injection isotope $^Z$A, maximal rigidity $R_{\mathrm{max}}$ and radius of the production region $r$. The studied ranges $10^9 \, \mathrm{GV} \lesssim R_{\mathrm{max}} \lesssim 5\, 10^{10} \, \mathrm{GV}$ and $5 \, 10^{16} \, \mathrm{cm} \lesssim r \lesssim 5 \, 10^{18} \, \mathrm{cm}$ are chosen to cover the parameters of the original study for all three TDEs, and to satisfy the constraints from the non-observation of electromagnetic cascades of hadronic origin \citep{2023ApJ...956...30Y}. The maximal rigidity, which earlier was an ad hoc choice beyond the photo-pion threshold of IR photons  \citep{2023ApJ...948...42W}, is varied in a reasonable range to potentially describe UHECR data, see e.g. \citet{2019ApJ...873...88H}.
The quasi-thermal photon temperatures, their densities, and their time evolutions are derived from observations of the individual TDEs, as described in detail in \citet{2023ApJ...948...42W}.   We fix the magnetic field to be $B\simeq 0.1 \, \mathrm{G}$ for the sake of simplicity, as in \citet{2023ApJ...948...42W}. A summary of the model's parameter values is provided in \tableref{model_parameters}. 

\begin{table}[t]
    \centering
    \begin{tabular}{ll|ccc}
    \hline
    \hline
     \textbf{Description} & \textbf{Symbol} & \multicolumn{3}{c}{\textbf{Value}} \\
     &  & {\bf AT2019dsg} & {\bf AT2019fdr} & {\bf AT2019aalc} \\
    \hline
    \multicolumn{5}{l}{\bf{Input parameters}} \\
    \hline

        \multirow{3}{*}{Photon temperature [eV] }& $k_BT_{\mathrm{IR}}$   &   0.16 (1) & 0.15 (2)&  0.16 (1) \\
        & $k_BT_{\mathrm{OUV}}$ at $t_{\mathrm{peak}}$  & 3.4 (3) & 1.2 (2) &  0.9 (1) \\
        & $k_BT_{\mathrm{X}}$  &  72 (3) & 56 (2,4)&   172 (4) \\
        \multirow{3}{*}{Photon luminosity [$\frac{\text{erg}}{\text{s}}$] } &$L_{\mathrm{IR}}^{\mathrm{bol}}$ at $ t - t_{\mathrm{peak}}$  
 & $2.8 \, 10^{43}$ at 431$\, \text{d}$  (1) & $5.2 \, 10^{44}$ at 277$\, \text{d}$   (1)  &  $1.1 \, 10^{44}$ at 123$\, \text{d}$ (1)  \\
        & $L_{\mathrm{OUV}}^{\mathrm{bol}}$ at $t_{\mathrm{peak}}$ & $2.8 \, 10^{44}$ (1) &  $1.4 \, 10^{45}$ (1) &   $2.7 \, 10^{44}$ (1) \\
        & $L_{\mathrm{X}}^{\mathrm{bol}}$ at $ t - t_{\mathrm{peak}}$ &  $6.2 \, 10^{43}$ at 17$\, \text{d}$ (3) & $6.4 \, 10^{43}$  at 609$\, \text{d}$  
        (2) & $1.6 \, 10^{42}$  at 495$\, \text{d}$ 
        (4)  \\
        Accretion rate factor at peak [$L_{\mathrm{edd}} / c^2$] & $F_{\mathrm{peak}}$ & $100$  & $100$  & $100$ \\
        Mass of the disrupted star [$M_\odot$] & $M_*$ & $0.6$  & $5.7$  & $6.3$ \\
        Dynamical timescale [days]   &$t_{\mathrm{dyn}}$  & $670$  & $1730$  & $1970$  \\
        Dissipation efficiency [$L_p/\left(\dot M c^2 \right)$] &$\varepsilon_{\mathrm{diss}}$  & $0.2$   & $0.2$ & $0.2$  \\
        Magnetic field [G] & $B$  & 0.1 &  0.1 & 0.1  \\
        CR maximal rigidity [GV]& $R_{\mathrm{max}}$ & $10^{9} - 5\, 10^{10}$ & $10^{9} - 5\, 10^{10}$ & $10^{9} - 5\, 10^{10}$\\
        Radius of the production region [cm] & $r$  &  $5\,10^{16} - 5\, 10^{18}$ &  $5\,10^{16} - 5\, 10^{18}$  &  $5\,10^{16} - 5\, 10^{18}$  \\
    \hline
    \multicolumn{4}{l}{\bf{Characteristics}} \\ 
    \hline
        Redshift &$z$ &  0.051 (3) & 0.267 (2) & 0.036 (4) \\
        SMBH mass  [$M_\odot$] & $M$ & $5.0_{-2}^{+16} \, 10^6$ (4) & $1.3 _{-0.4}^{+4.1} \, 10^7$ (4) & $1.6 _{-0.5}^{+5.1} \, 10^7$ (4) \\
        Expected number of neutrinos &$N_\nu$ &  0.008--0.76 (3) &  0.007--0.13 (2) & not available \\
        Radius of the IR region  [cm] &$r_{\mathrm{IR}}$ & $5\, 10^{18}$ (3) &   $5\, 10^{17}$ (2)& $1.6\, 10^{18}$ (4) \\
    \hline
    \end{tabular}
    \caption{ \textbf{Summary of the input parameters and characteristics for individual TDE model}.  The table is adapted from \citet{2023ApJ...948...42W}. References to the original articles are given in brackets: 
    (1) \citet{2023ApJ...948...42W}
    (2) \citet{Reusch_2022},
    (3) \citet{2021NatAs...5..510S}
    (4) \citet{van_Velzen_2021}. 
    The luminosities and photon temperature are provided at the indicated time; for details, see \citet{2023ApJ...948...42W}.}
    \label{tab:model_parameters}
\end{table}

\subsection{UHECR injection in TDEs}
\label{subsec:TDE_UHECR}

We extend the model of \citet{2023ApJ...948...42W} to nuclei following the methods in \citet{2017NatSR...7.4882B,2018A&A...611A.101B,2018NatSR...810828B}, which we refer to for details. We simulate the source for different injection isotopes, which disintegrate and interact with the IR, OUV, and X-ray target photons in a test-particle approach, leading to a nuclear cascade in the source. We perform the computation, including the full-time evolution of the target photons, which we also show below; for the main results, we use time-integrated neutrino and UHECR spectra since (at least for UHECRs) the time dependence cannot be measured. We then superimpose the results from different nuclei in the source-propagation chain, which is a possible simplification, since we do not take into account non-linear feedback (e.g., disintegration photons adding to the target). 

Our time-dependent solver is based on NeuCosmA, constructing a coupled Partial Differential Equation (PDE) system for all nuclear isotopes considered, protons, neutrons, muons, pions, kaons, and neutrinos, and evolving it with a Crank-Nicolson solver~\citep{1947PCPS...43...50C} of the PDE system re-parameterized on a double-log scale in $E$-$E^2 dN/dE$. The setup of the nuclear system is performed fully automatically with a recursive algorithm following all possible disintegration and beta decay channels beyond a chosen threshold value for the secondary multiplicity (1\%) -- which acts as a control parameter for the completeness of the system and for the precision of the results. Therefore, the size of the PDE system critically depends on the injection mass number $A$, and the computation time scales correspondingly. Radiation processes taken into account are nuclear disintegration, spontaneous (nuclei) emissions, beta decays, synchrotron cooling losses, photo-pion production, Bethe Heitler pair production losses, escape (discussed below) and $Ap$ interaction losses, whereas we assume a stable production region ($r \simeq const$), which means that adiabatic losses can be neglected. As a computational example 
a single run for AT2019fdr with $^{56}$Fe injection (largest $A$ considered) takes about two hours on a single workgroup server core. Out of 481 possible tabulated isotopes and 39962 tabulated disintegration channels (based on TALYS \citep{Koning:2012zqy}), 218 isotopes and 3838 disintegration channels are automatically selected in that case and evolved over the duration of the TDE. The same computation for proton injection takes only 30 seconds. 

\begin{figure*}[t]
    \begin{center}
    \includegraphics[width=.75\textwidth]{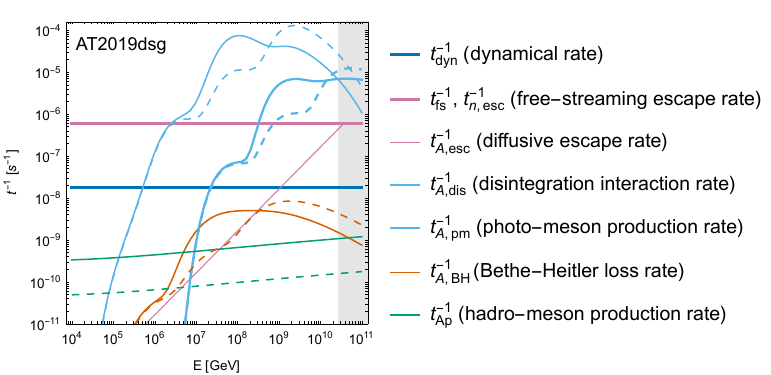}
    \end{center}
    \caption{Relevant interaction and energy loss rates (in the source frame) as a function of $E$ (in the observer's frame) for AT2019dsg and $^{Z}A=^{56}$Fe injection. Solid curves refer to the peak time of the TDE, and dashed curves refer to the time of the neutrino emission. The dynamical rate ($t_{\rm dyn}^{-1}$)} corresponds to the evolution of the whole system, and the free-streaming rate is $R/c$. The shaded region is beyond the assumed maximal injection energy.
    \label{fig:timesdsg}
\end{figure*}

For illustration in this section, we follow the original models in \citet{2023ApJ...948...42W} as closely as possible, which means that we use the same fiducial TDE parameters and also include an outflow with $v=0.5 \, c$ the nuclei may interact with. Here we inject $^{56}$Fe with an {\em ad hoc} choice $R_{\mathrm{Fe},\mathrm{max}} = 3.5 \, 10^{10} \, \mathrm{GV}$ (whereas later $R_{\mathrm{max}}$ will be varied). The relevant interaction and energy loss rates for the primary nucleus are shown in \figu{timesdsg} for AT2019dsg. What we see here is that for iron, nuclear disintegration dominates at the highest energies (thin blue curves), the source is optically thick, and the nuclear cascade will efficiently develop. Both the leading disintegration  (thin blue) and photo-meson (thick blue, leading to neutrino production) exhibit a triple-hump structure from interactions off (from lowest to highest) the X-ray, OUV, and IR target photons. Note that the solid curves refer to the peak time ($t_{\rm peak}$) of the TDE, and dashed curves refer to the time of the neutrino emission; since the IR dust echo is delayed with respect to the blackbody peak, disintegration (and photo-meson production) of the IR photons dominates at the time of the neutrino emission. Other processes are subleading at first but can become relevant for lighter nuclei over the evolution of the nuclear cascade as disintegration and photo-meson processes lead to nuclei with lower $A$, see \citet{2018A&A...611A.101B} for details. Our simulation includes this evolution self-consistently in a fully time-dependent computation.

\begin{figure*}[t]
    \begin{center}
    \includegraphics[width=.32\textwidth]{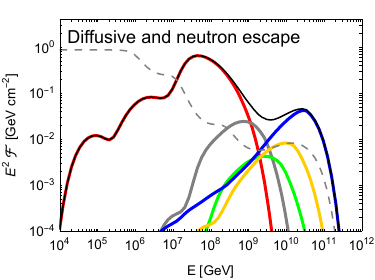}
    \includegraphics[width=.32\textwidth]{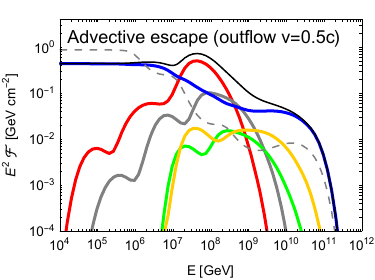} 
    \includegraphics[width=.32\textwidth]{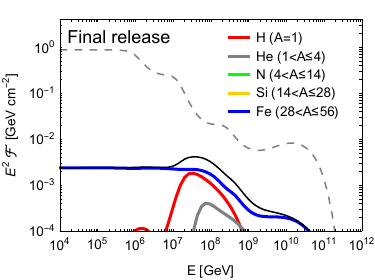} 
    \end{center}
    \caption{Different possible (time-integrated) cosmic-ray escape contributions for AT2019dsg (pure $^{56}$Fe injection; observer's frame with adiabatic losses during transport only). Left panel: diffusive (direct) escape including free-streaming neutron escape; middle panel: advective escape for the optimistic case of an outflow with $v=0.5 \, c$; right panel: final release at the end of the simulated TDE evolution (after $t_{\mathrm{dyn}}$), assuming that the injection ceases and the magnetic fields decay and all leftover non-thermal CRs are released. The different colors correspond to spectra summed over different mass groups, see legend in right panel: H (red, $A=1$), He (grey, $1 < A \le 4$), N (green, $4 < A \le 14$), Si (yellow, $14 < A \le 28$), and Fe (blue, $28 < A \le 56$), and the black curves show the total escape spectrum. Dashed curves illustrate the hypothetical spectrum if the primary $^{56}$Fe was escaping free-streamingly for comparison (which can be lower than the other curves because it represents a single isotope only).}
    \label{fig:escape}
\end{figure*}

It is an open question how UHECRs escape from the source, where ``escape spectra'' refer to the cosmic ray emission spectra ejected into the circum-source environment and injected into the UHECR propagation code, which describes transport through cosmic microwave background and cosmic radiation fields. In the present scenario, CRs at lower energies are confined in the production region $r$ by turbulent magnetic fields, whereas the electrically neutral neutrons can free-stream out of the production region -- unless it is optically thick to $n\gamma$ interactions. Since UHECRs can escape from the region if the Larmor radius $R_L=E_A/(ZeB)>r$, a suitable parameterization is given by \citep{2013ApJ...768..186B}
\begin{equation}
 t_{p,\mathrm{esc}}^{-1} = \min \left( t_{\mathrm{fs}}^{-1}  ,  \frac{D}{c \,  t_{\mathrm{fs}}^2} \right)  \, , \label{equ:pescape}
\end{equation}
with the diffusion coefficient $D \simeq R_L c$ in the Bohm-like regime and the free-streaming timescale $t_{\mathrm{fs}}=r/c$. The first term ensures that the escape is limited by the free-streaming (ballistic) rate. While a different diffusion coefficient leads to a slightly different result, and also the coherence length of the magnetic field affects the shape of the escape component (see, e.g. \citet{2022Physi...4..473B}), this approach describes relatively hard escape spectra required from UHECR data, see, e.g., \citet{2019ApJ...873...88H}. We show in \figu{escape}, left panel, the corresponding cosmic-ray escape spectra group into different mass groups (see figure caption) for AT2019dsg  and pure $^{56}$Fe injection without transport effects (other than adiabatic losses). It is clear from the figure that the spectral shape below the peak of each component is quickly masked by a different component, which implies that, for hard enough escape spectra, the fit will not be very sensitive to the precise spectral index, and thus to the details of the diffusion model there. The H mass group is driven by protons and neutrons produced in the disintegration, which is suppressed by photo-meson production at high energies. The $^{56}$Fe (hypothetical) free-streaming escape spectrum is shown as the dashed curve, which corresponds to the in-source density suppressed by disintegration losses. It is lower than the Fe curve because it represents a single isotope only, and disintegration is very efficient, as can be seen from the shape of the curve and the comparison with low energies, where disintegration does not occur. In fact, the escape is relatively inefficient compared to nuclear disintegration and photo-meson production, as can be seen in \figu{timesdsg}.

One could also think about other contributions to UHECR escape. For example, the outflow may transport particles away by advection, which leads to softer escape spectra following the in-source density, see middle panel of  \figu{escape}. This contribution relies on the presence of the outflow and depends on its velocity. Another contribution would be the final release of all confined nuclei when the system (such as magnetic fields) decay, which we illustrate in the right panel of \figu{escape}. This contribution is, however, small because most of the energy of the system has been dissipated otherwise. Since it turns out that the advective contribution cannot describe UHECR data very well, and the final release contribution is small, we further adopt the diffusive contribution only.\footnote{For self-consistency, we nevertheless include the effects of the outflow in terms of $pp$ interactions and an advective escape rate, which is similar to the effect of an adiabatic cooling rate. This leads to slightly smaller predicted neutrino fluences compared to \citet{2023ApJ...948...42W} (where advective escape/adiabatic cooling was not considered) and to slightly harder proton spectra at low energies (where the protons cannot pile up so much anymore). } While we only show the (summed) results over different mass groups in the figures in this section, we transfer all escape spectra of all isotopes in the system to the UHECR transport code at the interface (see below).
 
\begin{figure*}[t]
    \begin{center}
    \includegraphics[width=.45\textwidth]{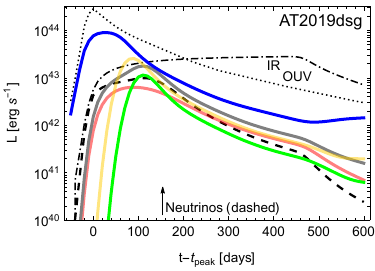}
    \includegraphics[width=.45\textwidth]{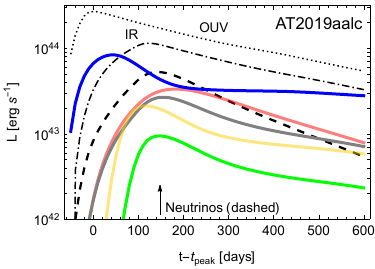}
    \end{center}
    \caption{Time evolution of the diffusive UHECR escape (luminosity) for the different mass groups  (same colors as in \figu{escape}) for AT2019dsg (left panel) and AT2019aalc (right panel). The  OUV (black dotted curves), IR (black dashed-dotted curves), and neutrino (black dashed curves) luminosities are shown as well ($\nu_\mu$ and $\bar\nu_\mu$ including flavor mixing); the arrows mark the actual times of the neutrino arrivals. Here, pure $^{56}$Fe injection is assumed, and the UHECR luminosities are integrated in the energy range above $10^9 \, \mathrm{GV}$ (observer's frame), whereas the neutrino luminosities are integrated over the full energy range. }
    \label{fig:evolution}
\end{figure*}

Even though we perform a fully time-dependent source simulation, we will, for most of this study, integrate the escaping cosmic-ray spectra over time because time-dependent effects for UHECRs are not measurable. However, the time-dependent escape is theoretically interesting, and the time-dependent spectra can have an effect on the neutrino production time. We, therefore, show in \figu{evolution} the time evolution of the diffusive UHECR escape luminosity for the different mass groups (colors) for AT2019dsg (left) and AT2019aalc (right), in comparison to the OUV and IR, and neutrino luminosities (black curves). The absolute luminosities should be interpreted with care because the UHECR luminosities are integrated in the energy range above $10^9 \, \mathrm{GV}$ (observer's frame), and therefore effects at low energies are not captured. It is interesting that the primary nuclei disintegrate efficiently in the beginning, leading to the time-delayed (by the disintegration loss rate) appearance of lighter nuclei. There also can be a late-time recovery of UHECR escape because the target photon fields decay and the injected UHECRs (following the OUV luminosity) can escape; this contribution is, however, small compared to the escape luminosity closer to the peak. Finally, note that, in the shown examples, the neutrino luminosities (dashed curves) peak close to the actual neutrino observation times better than for the proton model in \citet{2023ApJ...948...42W}. Here, the neutrino production is driven by the protons and neutrons at lower energies (see \figu{escape} and also \citet{2018A&A...611A.101B}), which are abundantly produced by the nuclear cascade and delayed by the disintegration cooling timescale. 

\subsection{Expected composition of accelerated cosmic rays from progenitors}
\label{sub:composition}

In order to accurately replicate UHECR data, a significant portion of the accelerated nuclei must be predominantly from the nitrogen and silicon groups, i.e., have a medium to medium-heavy composition \citep{Aab:2017,2019ApJ...873...88H, Plotko_2023}.  It is most natural to assume that these nuclei originate from the disrupted star and to identify their nuclear composition with the composition of the progenitor. That direct relationship is expected if there are no selection effects at play (e.g., favoring the acceleration of certain nuclear species over others) and if the black hole tidal forces do not change the composition of the stellar matter.  
Here, we consider several representative nuclear compositions motivated by the different types of disrupted stars. In particular, we examine main sequence stars (MS), ONeMg White Dwarfs (ONeMg-WD), CO White Dwarfs (CO-WD), Wolf-Rayet stars (WR), and red supergiants (RSG) (with composition averaged over their post-main-sequence lifetime), see \tableref{comptable}. MS, RSG, and WR stars have similar mass abundances, primarily consisting of H and He with small amounts of medium and heavy elements. On the other hand, CO and ONeMg-WD stars have mass abundances dominated by medium and heavy elements, making them historically more suitable for UHECRs.

In this study, we chose the injected nuclear composition motivated by stellar evolution \citep{leblanc2010introduction}: H, He, C, N, O, Na, Si, and Fe. Due to computation time restrictions and minimal impact on the final results, we couldn't include all possible elements, so the fractions of elements not selected were added to the closest nuclear isotope. For the composition of the injected CR ($f_i$), we consider two options. The first option involves a composition of injected CRs matching the mass abundance. This assumption holds under specific conditions, such as having the same minimal Lorentz factor $\gamma_{\mathrm{min}}$ for all particles and a rigidity-dependent maximum energy ($E_{i,\mathrm{max}}= Z_{i} eR_{\mathrm{max}}$). In contrast, the second option treats the injected CR composition as a free parameter and compares it to the mass abundance of a chosen set of stars.

\begin{deluxetable*}{l|r|r|r||r|r}[t]
\label{tab:comptable}
    \centering
    \caption {\textbf{Expected mass abundances for different types of stars:} Main Sequence ~\citep[MS,][]{Zhang:2017hom}, Red Super Giant ~\citep[RSG,][]{RSG_comp}, Wolf-Rayet ~\citep[WR,][]{WR_comp}, CO White Dwarf ~\citep[CO-WD,][]{Zhang:2017hom}, and ONeMg White Dwarf ~\citep[ONeMg-WD,][]{Zhang:2017hom}.}
    \tablehead{
        Mass
        & \multicolumn{3}{c||}{Star's stage}
        & \multicolumn{2}{c}{White Dwarf}
        \\ 
        abundance (\%)
        & \multicolumn{1}{c|}{MS}
        & \multicolumn{1}{c|}{RSG}
        & \multicolumn{1}{c||}{WR}
        & \multicolumn{1}{c|}{CO-WD}
        & \multicolumn{1}{c}{ONeMg-WD}
    }
    \startdata
    \multicolumn{1}{l|}{H}
        & 73.90
        & 46.46
        & 0.01
        & 0
        & 0
    \\
    \multicolumn{1}{l|}{He}
        & 24.70
        & 36.74
        & 98.10
        & 0
        & 0
    \\
    \hline
    \multicolumn{1}{l|}{C}
        & 0.22
        & 0.95
        & 0.03
        & 50 
        & 0
    \\
    \multicolumn{1}{l|}{N}
        & 0.01
        & 0.30
        & 1.33
        & 0
        & 0
    \\
    \multicolumn{1}{l|}{O}
        & 0.63
        & 2.72
        & 0.03
        & 50 
        & 12
    \\
    \hline
    \multicolumn{1}{l|}{Na}
        & 0.23
        & 0.99
        & 0.26
        & 0
        & 88
    \\
    \multicolumn{1}{l|}{Si}
        & 0.0
        & 0.30
        & 0.07
        & 0
        & 0
    \\
    \hline
    \multicolumn{1}{l|}{Fe}
        & 0.12
        & 0.52
        & 0.14
        & 0
        & 0
     \\
    \enddata
    
\end{deluxetable*}

\subsection{Tidal disruption constraints on SMBH mass from different progenitors}

A star is tidally disrupted when its tidal disruption radius $r_{\text{TDE}}= R_* \left(M / M_* \right)^{1/3} $ exceeds the Schwarzschild radius $r_s = 2GM/c^2$ of the black hole; here $M$ is the SMBH mass, $R_*$ is the star's radius, $M_*$ is the star's mass and $G$ is the gravitational constant. If \( r_{\text{TDE}} < r_s \), the star will cross the event horizon before being torn apart, preventing the formation of a luminous TDE. The maximum SMBH mass for which a star of given mass and radius can be disrupted is given by \citet{2009ApJ...695..404R}:
\begin{equation}
    M_{\text{max}} = 1.1 \times 10^8 \left(\frac{R_*}{R_\odot}\right)^{3/2} \left(\frac{M_*}{ M_\odot}\right)^{-1/2} M_\odot.
\end{equation}

\begin{deluxetable*}{l|l|c|c|c||cc}
\tablecaption{\textbf{Maximum SMBH masses} for tidal disruption of various progenitor types: Main Sequence ~\citep[MS,][]{Zhang:2017hom}, Red Super Giant ~\citep[RSG,][]{RSG_comp}, Wolf-Rayet ~\citep[WR,][]{WR_comp}, CO White Dwarf ~\citep[CO-WD,][]{Zhang:2017hom}, and ONeMg White Dwarf ~\citep[ONeMg-WD,][]{Zhang:2017hom}. The assumptions for $M_*$ and $R_*$ are listed in the table.}
\label{tab:SMBH_limits}
\tablehead{
    Description & Symbol
    & \multicolumn{3}{c||}{Star's stage}
    & \multicolumn{2}{c}{White Dwarf}
    \\ 
    &  & MS & RSG & WR & CO-WD & ONeMg-WD
}
\startdata
Stellar mass [$M_\odot$] & $ M_*$ 
    & 1.0 
    & \( 1.2 \, 10^{1} \)
    & 7.5
    & 0.6
    & 1.1 \\
Stellar radius [$R_\odot$] & $ R_* $ 
    & 1.0 
    & \( 5.0 \, 10^{2} \)
    & 3.6 
    & \( 1.0 \, 10^{-2} \)
    & \( 1.0 \, 10^{-2} \) \\
\hline
Upper limit on SMBH mass [$M_\odot$] & $M_{\text{max}}$ 
    & \( 1.1 \, 10^8  \)
    & \( 3.6 \, 10^{11} \)
    & \( 2.7 \, 10^8 \)
    & \( 1.4 \, 10^5\)
    & \( 1.0 \, 10^5 \) \\
\enddata
\end{deluxetable*}

\tableref{SMBH_limits} lists the upper limits on SMBH mass for different progenitor types. The SMBHs associated with AT2019dsg, AT2019fdr, and AT2019aalc have estimated masses around \( 10^7 M_\odot \), indicating that main sequence stars, red supergiants, and Wolf-Rayet stars can be tidally disrupted in these events. However, white dwarfs (both CO-WD and ONeMg-WD) have much lower \( M_{\text{max}} \) values, suggesting that they would likely plunge directly into the black hole before experiencing tidal disruption.

These calculations assume a non-rotating (Schwarzschild) black hole, whereas  for of a spinning (Kerr) black hole, the disruption conditions are modified. A rapidly rotating SMBH increases the tidal disruption radius relative to the event horizon, effectively increasing the maximum mass for which disruption can occur. For a maximally spinning Kerr black hole, \( M_{\text{max}} \) can be up to 1.5-10 times higher than in the non-spinning case \citep{1994MmSAI..65.1135S, Kobayashi_2004, Mummery_2023}. This means that white dwarfs that would otherwise be swallowed as a whole by a non-rotating SMBH could still undergo partial disruption near a rapidly spinning SMBH.

While the direct disruption of white dwarfs in AT2019dsg, AT2019fdr, and AT2019aalc appears highly unlikely, even when considering spinning effects and uncertainties in the SMBH masses associated with these events, we include WD-like compositions in our modeling for comparison. This is motivated by previous studies \citep{Zhang:2017hom,2018NatSR...810828B}, which indicate that a white dwarf-like composition are necessary to explain the observed cosmic-ray composition at Earth. Although our primary focus is on fitting the composition to the available data, with main-sequence stars as the dominant progenitors, it is important to emphasize this clear tension between the high SMBH masses, which strongly disfavor white dwarf disruptions in these events, and the heavy-element composition inferred from UHECR observation.

\subsection{Consistency with individual neutrino-TDE associations}

It's important to validate the consistency of our individual TDE models with the observed data before studying the UHECRs from a population of TDEs. The key observation is the expected number of muon neutrinos for three specific TDEs/candidates: AT2019aalc, AT2019fdr, and AT2019fdsg (see \tableref{model_parameters}). Our model involves three free parameters: the radius of the production region $r$, the maximum rigidity $R_{\mathrm{max}}$, and the composition of injected CRs. It's crucial to identify the allowed regions of these parameters that align with the expected number of muon neutrinos.

\begin{figure*}[htpb!]
    \centering
    \includegraphics[width=.7\textwidth]{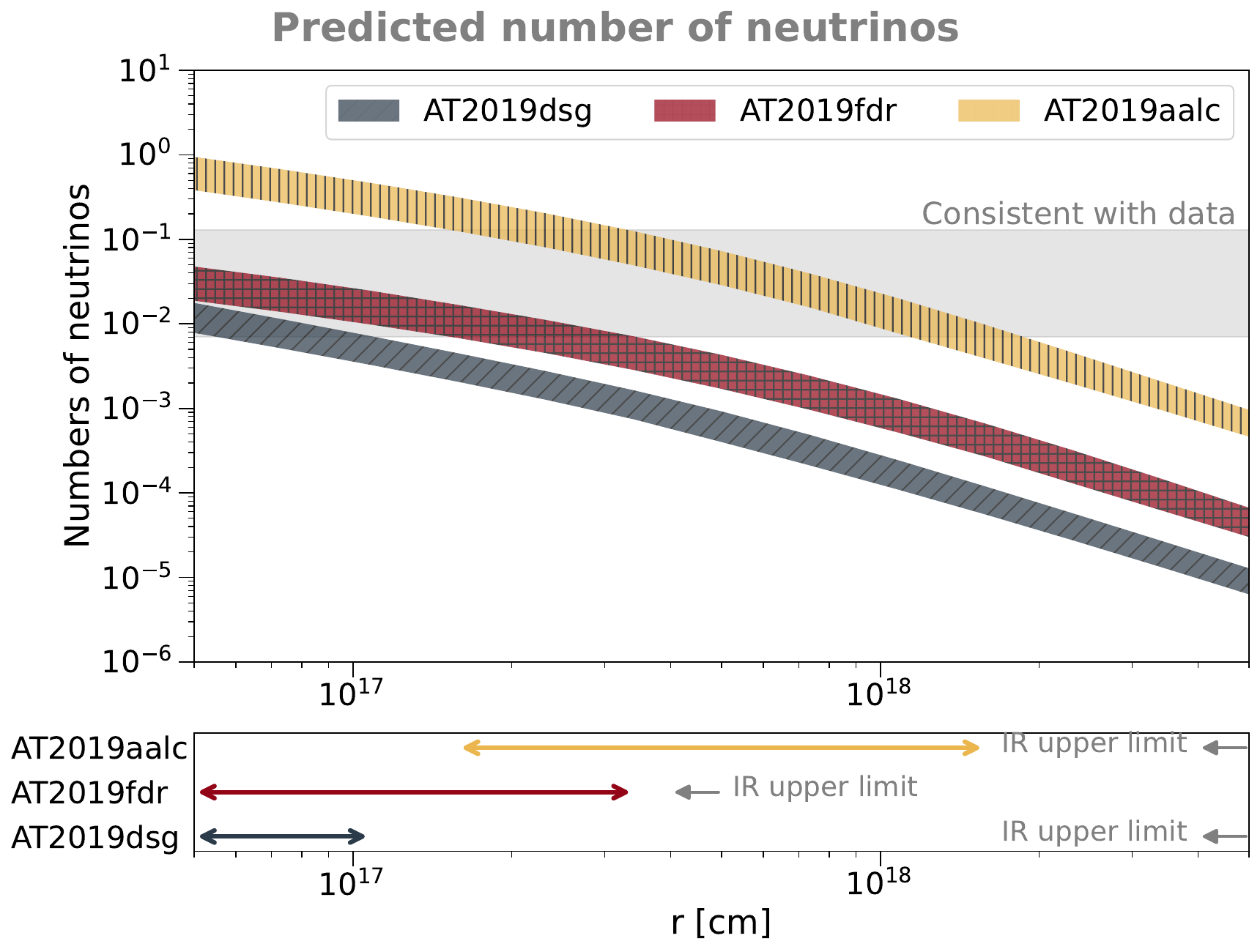} 
    \caption{\textit{Top:} The predicted number of muon neutrinos as a function of the production region radius for three TDEs/candidates: AT2019dsg, AT2019fdr, and AT2019aalc. The thickness of each curve accounts for maximum rigidity and cosmic-ray composition. The minimum and maximum numbers of neutrinos for each radius value are calculated by scanning over the possible range of the maximum rigidity and CR injection. \textit{Bottom:} The allowed region of the production region radius for all three TDEs based on the expected number of muon neutrinos from the top plot. Grey arrows mark the upper limits on the production region's radius derived from IR observations (see \tableref{model_parameters} and corresponding references). }
    \label{fig:N_nu_progenitors}
\end{figure*}

We start by calculating the muon neutrino fluence at Earth \(\mathcal{F}_{\nu_\mu}(E_\nu)\) for each TDE across a grid for both the radius of the production region ($r$) and the maximum rigidity ($R_{\mathrm{max}}$), with the radius ranging from \(5 \times 10^{16}\) to \(5 \times 10^{18}\) cm and the maximum rigidity ranging from \(10^9\) to \(5 \times 10^{10}\) GV. We consider pure injection for CRs of a single isotope from the seven possible injection isotopes (H, He, C, N, O, Na, Si, Fe). The next ingredient is effective area \( A_{\mathrm{eff}}(E_\nu)\), which represents the IceCube detector's sensitivity to neutrinos across different energy ranges. We utilize the Gamma-Ray Follow-up (GFU) effective area as described in \citep{Blaufuss:20199c}. The predicted number of muon neutrino events \(N_{\nu_\mu}\) by a model is then calculated by combining the predicted fluences with the effective areas over the entire energy range. This calculation can be expressed as:
\begin{equation}
N_{\nu_\mu} = \int \mathcal{F}_{\nu_\mu}(E_\nu) \, A_{\mathrm{eff}}(E_\nu) \, d E_\nu \label{equ:aeff}
\end{equation}
The next step involves comparing the predicted number of muon neutrinos expected by IceCube. Based on the previous studies, there are certain expectations for the number of muon neutrinos from AT2019dsg and AT2019fdr, but such expectations are not available for AT2019aalc. For instance, the detection of a single high-energy neutrino associated with AT2019dsg suggests a mean expectation in the range of 0.008 to 0.76 muon neutrinos at the 90\% confidence level, considering the total TDE population observed by the Zwicky Transient Facility (ZTF) \citep{2021NatAs...5..510S}. Similarly, for AT2019fdr, the expected number of muon neutrinos is between 0.007 and 0.13, based on its contribution to the g-band peak energy flux among nuclear transients \citep{Reusch_2022}. Based on this information, we use the allowed range of 0.007-0.13 muon neutrinos for our study.  As indicated in the study by \citet{2023ApJ...948...42W}, the radius of the production region is the main control parameter for the expected number of neutrinos. Consequently, we present our findings graphically against the radius as shown in \figu{N_nu_progenitors}. For each value of the radius of the production region, we calculate the minimum and maximum numbers of neutrinos across the entire range of maximum rigidity and all injected elements.

The expected number of neutrinos generally depends on four fixed parameters in our model: the size of the production area, the bolometric IR luminosity, and the redshift. AT2019fdr is the farthest from us \( z =0.267\), but its bolometric IR luminosity is the highest among the three TDEs ($L_{\mathrm{IR}}^{\mathrm{bol}} \approx 5.2 \, 10^{44} \mathrm{erg}/\mathrm{s} $ at 227 days post-peak). For AT2019dsg, the luminosity is the lowest ($L_{\mathrm{IR}}^{\mathrm{bol}} \approx 2.8 \, 10^{43} \mathrm{erg}/\mathrm{s} $ at 431 days post-peak), but it is closer \( z = 0.051\), which leads AT2019fdr and AT2019dsg to predict a similar number of neutrinos for the same value of the radius. AT2019aalc has the best combination, being the closest \( z=0.036\) and having a high bolometric IR luminosity ($L_{\mathrm{IR}}^{\mathrm{bol}} \approx 1.1 \, 10^{44} \mathrm{erg}/\mathrm{s} $ at 123 days post-peak), making AT2019aalc the most efficient neutrino TDE among all three.

In terms of the allowed region for the radius,  the expected numbers of neutrinos for AT2019fdr and AT2019dsg, shown as the grey-blue and red curves in \figu{N_nu_progenitors}, fall within the allowed neutrino region (marked as a grey area on the plot) only at smaller radii. AT2019dsg shows agreement with expected values only at the smallest radii from $5\, 10^{16}$ to $1.6\, 10^{17}$ cm. AT2019fdr indicates a preference for smaller radii falling within the range of $5\, 10^{16}$ to $2.32\, 10^{17}$ cm. Since AT2019aalc is the most efficient neutrino TDE, the allowed region is shifted to a bigger radius value from $1.6\,10^{17}$ to $2.6\,10^{18}$ cm. Besides the restriction from the expected number of muon neutrinos, there is another independent method to estimate upper limits on the production radius involving IR observations. By interpreting the time-delayed IR emission as a dust echo, the size of the production region can be estimated (see \tableref{model_parameters} and corresponding references).  For all TDEs, the allowed region based on the expected number of neutrinos is inside the one based on IR data (grey arrows on the bottom plot in \figu{N_nu_progenitors}).

\begin{figure*}[htpb!]
    \centering
    \includegraphics[width=.65\textwidth]{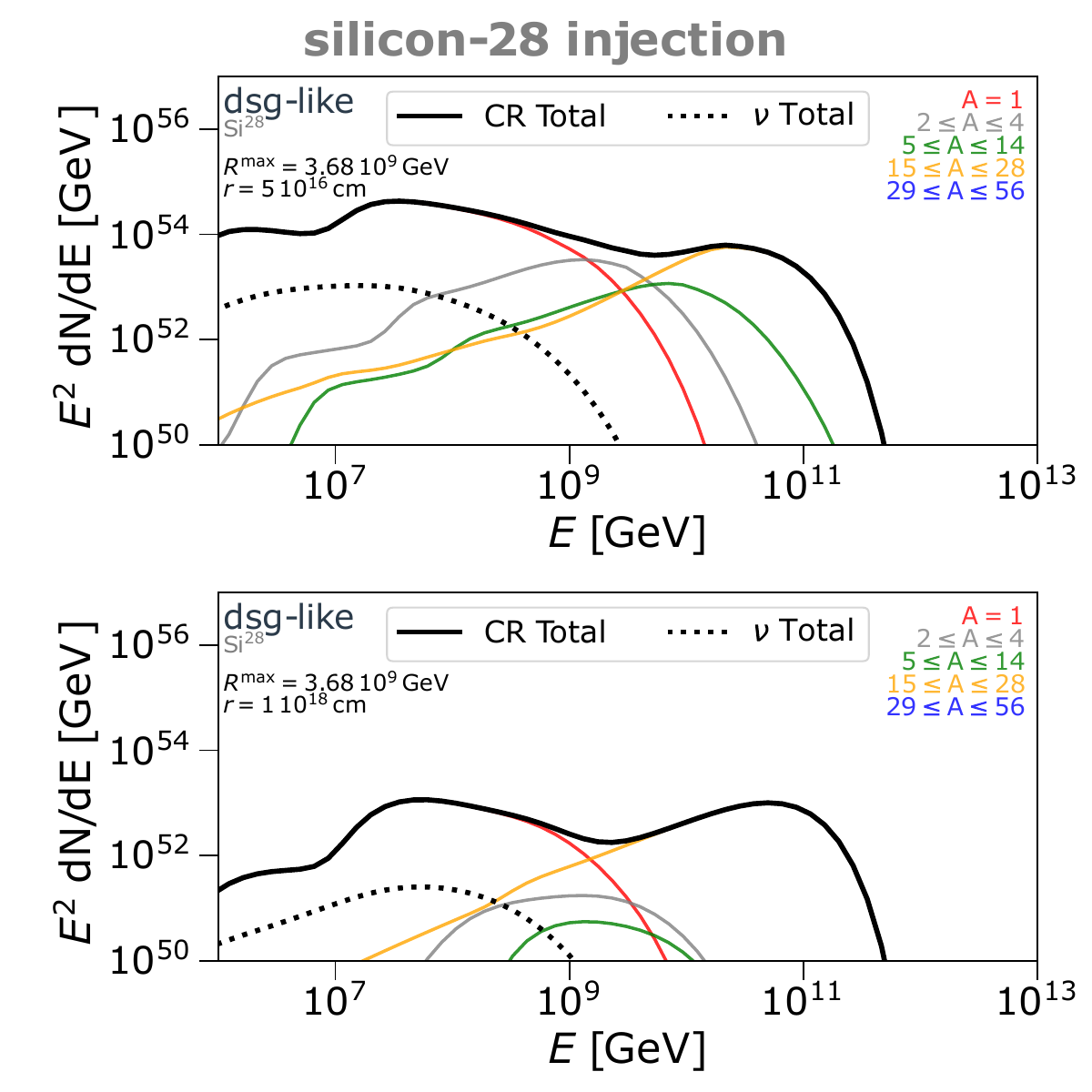} 
    \caption{The escaped spectra of CRs and neutrinos from an AT02219dsg-like event for two radii ($5 \, 10^{16}$ and $1 \, 10^{18}$ cm), under the same conditions of maximum rigidity ($3.68 \, 10^{9}$ GV) and the CR injection (pure silicon-28), integrated over the entire duration of each event. The solid curves represent the total escaped spectrum of CR, with the contribution of individual CR groups shown in different colors. The black dotted curve represents the total neutrino flux.}
    \label{fig:source_spectrum_dsg_two_radii}
\end{figure*}

Apart from influencing the expected number of neutrinos, the production radius also significantly affects the CR disintegration within the radiation zone. \figu{source_spectrum_dsg_two_radii} shows the escaped spectra of CRs and neutrinos for two distinct radii — $5 \, 10^{16}$ and $1 \, 10^{18}$ cm — while maintaining the same maximum rigidity ($3.68 \, 10^{9}$ GV) and CR injection (pure Si). A smaller radius increases the CR interactions with background photons before they can escape, leading to a higher flux of secondary particles at similar energies of the injected primary element. Conversely, a larger radius allows CR survival. As demonstrated by  ~\citet{Aab:2017} and  ~\citet{Heinze:2019}, UHECR data indicate sources with a peaky quasi-monochromatic spectrum of isotopes transitioning from protons at lower energies to iron at higher energies. 
For our study, this means that a ``contamination'' of secondary nuclei in the escaped spectrum from disintegration makes the source unsuitable as UHECR source. This especially affects AT2019dsg, for which fixing the radius to smaller values, as required to produce a high enough neutrino fluence, implies such disintegration. Consequently, we assume that the dsg-like TDEs do not significantly contribute to UHECR and exclude it from our analyses. The following section will focus only on AT2019aalc and AT2019fdr.

\clearpage

\section{Modeling of UHECR-neutrino emitting populations of TDEs}
\label{sec:TDE_population}

In this section, we will discuss the challenges associated with creating a model for a UHECR-neutrino emitting population of TDEs based on two specific models of TDEs: AT2019aalc and AT2019fdr. Our focus will be on numerically simulating the propagation of UHECRs and neutrinos from various options for a population of TDEs, in order to fit the observed spectrum and composition of UHECRs without exceeding the current predictions for the diffuse neutrino flux measured by IceCube.

\subsection{Black hole mass function and TDE rates}
\label{sub:tderates}

 For each class of TDE (i.e., TDEs involving MS stars, WD, etc.), we use the \emph{total} cosmological rate of TDEs (number of TDEs per comoving volume per unit time) as a function of $z$, of the form $\rho(z)=\rho_0 (1+z)^{-3}$. The redshift dependence reflects the (negative) cosmological evolution of the SMBH and IMBH mass function, as described in, e.g. \cite{Shankar:2007zg} (see also \cite{Stone:2014wxa,Kochanek:2016zzg,Lunardini:2016xwi}).
 The normalization, $\rho_0$, depends on the details of the galactic population of stars and of SMBH/IMBH, and in particular on the disruption rate per black hole, $\dot N_{\mathrm{TD}}$, and on the black hole occupation fraction, $f_{\mathrm{occ}}$ (the probability that a SMBH or IMBH is located at the center of a host galaxy).

For main sequence star disruptions, a major source of uncertainty is the interval of SMBH mass, $M$, for which tidal disruption is efficient: varying the lower mass cutoff $M_{\mathrm{min}}$ can cause an uncertainty of about an order of magnitude: $\rho_{\mathrm{0}} \sim (0.7 - 20) \, 10^{3}~\mathrm{ Gpc^{-3} yr^{-1}}$  (see \cite{Lunardini:2016xwi}), whereas the effect of the upper cutoff, $M_{\mathrm{max}}\sim 10^7 - 10^8 \msun$ -- the mass beyond which a star is swallowed whole and not disrupted ---  is smaller.  

Disruption of white dwarfs must be 
significantly rarer, due to several factors. One of these is the smaller tidal disruption radius: in Ref. \cite{Krolik:2011fb} it is reasoned that - if a similar number density of WD and MS stars is assumed - $\dot N_{\mathrm{TD}}$ is suppressed by the  ratio of the tidal disruption radii: $R^{\mathrm{WD}}_{*}/R^{\mathrm{MS}}_{*}\sim 10^{-3}$. 
Further suppression may come from the flattening of the BH mass function at $M\lesssim \text{few} ~10^{5}\msun$ (see discussion in \cite{MacLeod:2015jma} and recent results in \cite{2024ApJ...969...93C}), and from the possibly low (and uncertain) occupation fraction for IMBH \citep{2024ApJ...969...93C}. 
Considering the large uncertainties, rates in the range $\rho_0 \sim 0.1- 50 \, \mathrm{Gpc^{-3}} \, \mathrm{yr^{-1}}$, 
are realistic, see, \eg\ \cite{Krolik:2011fb,Shcherbakov:2012zt,Zhang:2017hom}. Extreme values, up to $\rho_0 \sim 500 \, \mathrm{Gpc^{-3}} \, \mathrm{yr^{-1}}$  are found using recent numerical simulations \citep{Tanikawa:2021zfm}. 

For post-main-sequence stars, rates are typically lower than for MS stars. For example, the rate is estimated at $\rho_{\mathrm{0}} \sim 1 \, \mathrm{Gpc^{-3}} \, \mathrm{yr^{-1}}$ for red giants \citep{Syer:1998xh}, and should be orders of magnitude smaller for red supergiants, which are a lot more rare (a specific rate estimate could not be found for these).

We stress that the estimates above are tentative at best and could change significantly as a more detailed description of the tidal disruption process  --  taking into account elements like binary star systems, partial tidal disruptions, secondary black holes away from the galaxy's center, black hole spin, etc. --  emerge in the future.

\subsection{From two to many: Extrapolating the model to the TDE population}
\label{sub:maptopopulation}
Considering the sparseness of observations from only two selected \tds, applying our model to the entire UHECR-neutrino emitting \td\ population is challenging. Rather than attempting to be comprehensive, here we will discuss some plausible, although necessarily simplistic, scenarios. Simplification is necessary to fix the ideas and keep the computation numerically tractable. 

A key question here is how to assign values of the fitted parameters and other relevant input parameters (e.g., IR photon temperature and luminosity, etc.) to each member of the UHECR-neutrino emitting \td\ population. We are taking a model-independent approach and assuming that each of the AT2019aalc and AT2019fdr represents its own unique group. We will refer to them as the aalc-group and the fdr-group, which together represent the total population. Each group has its unique parameters and local rate ($\rho_i$, where $i=$aalc or fdr). The local rate of the entire UHECR-neutrino emitting TDE population is given by $\rho_0 = \sum_{i} \rho_i$. The analyzed range and values of these parameters are summarized in  \tableref{parameters_group}

Our analysis works as follows: the \td\ population is divided into two groups of identical objects: \textbf{aalc- and fdr-groups}. Each group is described by fixed parameters that are determined by the data on the two selected \tds\, and several variable parameters that will be fitted to the observed \uh\ data. These parameters are the maximum rigidity ($R_{\mathrm{max}}$), the radius of the production region ($r$), local rate $\rho_i$,  and the composition of injected CRs ($f$). 

Considering the large number of fitting variables and the limited amount of data, we have to consider different scenarios, which are compared to each other. In particular, we examine the following:

\begin{enumerate}
    \item \textbf{Mono Scenario} (Dominance of a single TDE group). In this scenario, we assume that one of the TDE groups is dominant while the other group is suppressed. For example, the aalc-group dominates when $\rho_{\mathrm{fdr}} \approx 0$. This approach helps us ensure that we don't overfit the data with a more complex model and allows us to test the scenario where a predominantly single group can explain the observed UHECR data.
    \item \textbf{Democratic Scenario} (Same parameters for all TDE groups).  Here, all the fitting parameters are considered the same for the entire \td\ population. That represents the situation where the local rates of the groups are democratic (same $\rho_{\mathrm{aalc}} = \rho_{\mathrm{fdr}} = \rho_0/2$), all the disrupted stars belong to the same class (e.g., main sequence stars), and the UHECR production mechanisms are as similar across the population as the current data allow. In this scenario, the production region for the aalc- and fdr-groups overlaps only within a small region around $2.32 \times 10^{17}$ cm, based on constraints from neutrino data for AT2019aalc and AT2019fdr.
    \item \textbf{Diverse Scenario} (Different parameters for all TDE groups).  Here, we describe a more realistic situation where the TDE population has substantial diversity. We consider scenarios where the aalc- and -fdr groups have different fitting parameters. This allows for variations in physical conditions and disruption rates across different types of TDEs.
     
\end{enumerate}

For each scenario, we are considering two options for the composition of the injected CR, as discussed in Sec. \ref{sub:composition}. The first option follows the maximum abundance of a star from the selected statistics. The second option treats the composition of the injected CR as fitting parameters.

\begin{table}[t]
    \centering
    \begin{tabular}{l|cc}
    \hline
    \hline
    Fitting parameters & {\bf aalc-group} & {\bf fdr-group} \\
    \hline
        $R_{\mathrm{max}} [\mathrm{GV}]$ & \multicolumn{2}{c}{$10^{9} - 5\, 10^{10}$}\\
        $r [\mathrm{cm}]$  &  $2.32\, 10^{17} - 1.58\, 10^{18}$ & $5\, 10^{16} - 2.32\, 10^{17}$
        \\
        $\rho_i$ [$ \mathrm{Gpc^{-3} yr^{-1} }]$  & \multicolumn{2}{c}{Not restricted}\\
        Injected CR fraction $f$ [\%] & \multicolumn{2}{c}{ MS, CO-WD, NeMgO-WD, or Fitted}\\
        \hline
    \end{tabular}
    \caption{ \textbf{Summary of fitting parameters for TDE groups}; The possible range of parameters used in the fit for modeling the aalc- and fdr- groups. }
    \label{tab:parameters_group}
\end{table}

\subsection{UHECR transport to Earth and description of UHECR data}
\label{subsec:UHECR_prop}

In our analysis of UHECRs originating from TDEs, we employ the open-source software \prince~\citep{Heinze:2019}, designed to numerically solve the one-dimensional transport equations, to model the propagation of CRs in the Universe and compute the cumulative cosmogenic neutrino spectra. We propagate the CRs escaped from the source for each aalc- and fdr-group, considering varying radius, maximal rigidity, and injected composition. We consider a redshift evolution \( \propto \left(1+z\right)^{-3} \) up to $z=5$ for the TDE populations to calculate the cumulative diffuse neutrino fluxes. For the configuration of the calculation, we use {the build-in} \textsc{Talys}~\citep{Koning:2007} nuclear interaction model, and use the Extragalactic Background Light (EBL) model by~\citet{Gilmore:2012}. 

Our primary goal is to fit the UHECR spectrum \citep{Verzi:2019AO} and composition \citep{Yushkov:2019J8} data above ${E_\text{min} = 6\,10^9 ~ \text{GeV}}$, observed by the Pierre Auger Observatory (Auger). In addition, we ensure that the diffuse neutrino flux from the TDE population does not exceed the High-Energy Starting Events (HESE) data from IceCube. For the composition fitting, we attempt to reproduce the observed mean depth of shower maximum density (\MeanXmax) and its standard deviation (\SigmaXmax), using the EPOS-LHC air-shower model \citep{Pierog:2015}\footnote{We also tested the Sibyll 2.3d air-shower model \citep{PhysRevD.102.063002} and found no significant difference. The results from Sibyll 2.3d air-show model are available at \url{plotkopavlo.com/uhecrsources/tde}.}. To account for systematic, we include the energy and  \MeanXmax\, uncertainties. However, we exclude the \SigmaXmax\, uncertainties since they have a minor influence on our analysis.

To evaluate the goodness of the fitting, we adopt the method from \cite{Plotko_2023} and calculate the $\chi^2$ value for the corresponding model $j$ of the TDE population:
\begin{align}
    \chi_{j}^2= \chi_{j, \mathrm{spectrum}}^{2} + \chi_{j, \langle X_\mathrm{max}\rangle}^2  +\chi_{j, \sigma(X_\mathrm{max})}^2 + \chi_{j, \nu}^2+
    \left(\frac{\delta_E }{\sigma_{\text{E}}}\right)^2  + \left( \frac{\delta_{\langle X_\mathrm{max}\rangle}}{100\%}\right)^2.
\end{align}
To ensure that the modelled spectrum does not exceed the UHECR fluxes at low energies, we incorporate five data points in the energy range from $2\,10^9 ~ \text{GeV}$ to  $6\,10^9 ~ \text{GeV}$. The term $\chi_{j, \nu}^2$  is a penalty to ensure that our model does not exceed the HESE diffuse neutrino flux. We consider the values of $\delta_E$ and $\delta_{\langle X_\mathrm{max}\rangle}$, defined as the differences between the model predicted results and the observations, within the $3 \sigma$ uncertainty range of corresponding data set.

Further, to compare different models of the TDE population,  we compute the Akaike Information Criterion corrected ~\citep[$\mathrm{AICc}$,][]{Akaike:74,Kenneth:2004, Buchner:2014,Rosales:2020} through the given equation:
\begin{align}
    \mathrm{AICc}_{j} = \chi^2_{j} + 2k_j + \frac{2k_j^2+2k_j}{n-k_j-1}, 
\end{align}
where $n$ is the number of data points, and $k_j$ is the number of parameters of the corresponding model $j$. The model with the smallest $\mathrm{AICc}$ value explains the data best among tested models. It is important to highlight that $\mathrm{AICc}$ should not be used as a null-hypothesis test~\citep{Anderson2000, Mundry2021}. In our study, we compare the $\mathrm{AICc}$ score of each model against  $\mathrm{AICc}$ of the model with the smallest value ($\mathrm{AICc}_{\mathrm{min}}$), calculating the difference ($\Delta_i$) as:
\begin{align}
    \Delta_{j} = \mathrm{AICc}_{j} - \mathrm{AICc}_{\mathrm{min}}. 
\end{align}
To make it easier to observe the differences between models, we compute $\Delta_j$ in the units of standard deviations following \citet{Plotko_2023}, and use it as an identifier for model selections.

\section{Results: fitting TDE population models to UHECR and neutrino Data}

\subsection{Evaluation of TDE population models}
\label{subsec:TDE_population_models}

We evaluate various configurations of UHECR-emitting TDE populations to identify models that accurately describe the observed UHECR spectrum and composition distributions without exceeding the constraints of IceCube's diffuse neutrino flux. The statistical comparison of these models is presented in \tableref{result_models_goodness_epos}, including $\chi^2$ values, degrees of freedom, and AICc values.

As expected, the Diverse Scenario best explains the data among the considered scenarios. In this scenario, the aalc- and fdr-groups have distinct fitted parameters, allowing for a better match to the observed UHECR spectrum and composition. The Democratic Scenario, where all groups share the same parameters, is excluded by more than 8$\sigma$, indicating that uniform fitted parameters across all groups cannot effectively explain the UHECR data. The Mono Scenario, which assumes only one dominant group while excluding the others, is only viable when the aalc-group is dominant, but it is still excluded by $2.7\,\sigma$. These findings suggest that at least two distinct groups with different fitted parameters are necessary to reproduce the observed data. Hence, we examine the Diverse Scenario more closely in the following subsections.

In cases where the composition of injected CRs is based on the star mass abundances, the performance is worse compared to the Fitted Composition Case, where the composition of each nuclei group is allowed to vary freely. The Diverse Scenario performs the best when the TDEs from the aalc-group disrupt the ONeMg-WD, and the TDEs from the fdr-group disrupt the WR. Although, this case is still excluded by 4.1$\sigma$ compared to the Fitted Composition Case. Based on these findings, we present results only for the Fitted Composition Case of the Diverse Scenario. 

\begin{deluxetable*}{l||r|r||r|r||r|r|r}[tp]
    \centering
    \caption{\textbf{Comparison of fit results for different models of UHECR-neutrino emitting TDE population.} The statistical comparison of various TDE population models based on their fit to the observed UHECR spectrum and composition, as well as their consistency with IceCube's diffuse neutrino flux constraints. For each model, the table lists the assumption for the composition of injected CRs and the local rate for the aalc- and fdr-groups. For fit quality metrics, we show the $\chi^2$ value per degree of freedom (d.o.f.), the difference in Akaike Information Criterion corrected values ($\Delta \mathrm{AICc}$) with the best performing model, and the number of standard deviations ($\sigma$) by which each model is excluded relative to the best-fitting model. Models that are excluded by more than 8$\sigma$ are shown in grey. The best-fitting model is highlighted in bold. The WD compositions are shown for comparison only; WD disruptions are unlikely for AT2019dsg, AT2019fdr, or AT2019aalc, see \Sec~\ref{sec:discussion}}.
    \label{tab:result_models_goodness_epos}
    \tablehead{
        \textbf{Scenario}
        & \multicolumn{2}{c||}{\textbf{Composition}}
        & \multicolumn{2}{c||}{\textbf{ Group's local rate}}
        & \multicolumn{3}{c}{\textbf{Fit quality metrics}} \\
        &  \multicolumn{2}{c||}{ }
        & \multicolumn{2}{c||}{$\rho_i ~\rm[ Gpc^{-3} yr^{-1}]$}
        &  \multicolumn{3}{c }{ $\,$}\\
        & aalc-group
        & fdr-group
        & aalc-group
        & fdr-group
        & $\chi^2$ / d.o.f.
        & $\Delta \mathrm{AICc}$
        & $\sigma$
    }
    \startdata
    \multicolumn{1}{l}{Mono}\\
    \hline
      \multirow{4}{*}{$\,$$\,$$\,$$\,$$\,$$\,$$\,$$\,$$\,$$\,$$\,$$\,$$\,$$\,$$\,$$\,$ aalc}&\textcolor{gray}{MS} &  \multirow{4}{*}{ } & \textcolor{gray}{10} &  \multirow{4}{*}{ } & \textcolor{gray}{654/29} & \textcolor{gray}{603} & \textcolor{gray}{$>8$} \\
         &CO-WD & & 59 &   & 84/29 & 33 & 5.3 \\
         &\textcolor{gray}{ONeMg-WD} & & \textcolor{gray}{75} &   & \textcolor{gray}{192/29} & \textcolor{gray}{140} & \textcolor{gray}{$>8$} \\
         &Fitted & & 122 &   & 57/27 & 12 & 2.8 \\
    \hline
         \multicolumn{1}{r||}{fdr}& \textcolor{gray}{ } & \textcolor{gray}{Fitted} & \textcolor{gray}{ }  & \textcolor{gray}{9}  & \textcolor{gray}{261/27} & \textcolor{gray}{216} & \textcolor{gray}{$>8$ } \\
    \hline
    \multicolumn{1}{l}{\textcolor{gray}{Democratic}}\\
    \hline
        &\multicolumn{2}{c||}{\textcolor{gray}{MS}} & \multicolumn{2}{c||}{\textcolor{gray}{69}} & \textcolor{gray}{37650/29} & \textcolor{gray}{37599} & \multirow{4}{*}{\textcolor{gray}{$>8$ }} \\
        &\multicolumn{2}{c||}{\textcolor{gray}{CO-WD}} & \multicolumn{2}{c||}{\textcolor{gray}{75}} & \textcolor{gray}{2420/29} & \textcolor{gray}{2369} & \\
        &\multicolumn{2}{c||}{\textcolor{gray}{ONeMg-WD}} & \multicolumn{2}{c||}{\textcolor{gray}{69}} & \textcolor{gray}{447/29} & \textcolor{gray}{396} & \\
        &\multicolumn{2}{c||}{\textcolor{gray}{Fitted}} & \multicolumn{2}{c||}{\textcolor{gray}{69}} & \textcolor{gray}{248/27} & \textcolor{gray}{203} & \\
    \hline
    \multicolumn{1}{l}{\textbf{Diverse}}\\
    \hline
         &ONeMg-WD & WR & 47 & 5 & 57/24 & 23 & 4.3 \\
         &\textbf{Fitted} & \textbf{Fitted} & \textbf{194} & \textbf{13} & \textbf{38/25} & \multicolumn{2}{c}{\textbf{Best-fit case}} \\
    \hline
    \enddata
\end{deluxetable*}

\clearpage
\subsection{Describing the UHECR spectra and composition with TDE population}
\label{subsec:best_fit_scenarios}

\begin{figure*}[tp]
    \centering
    \includegraphics[width=.7\textwidth]{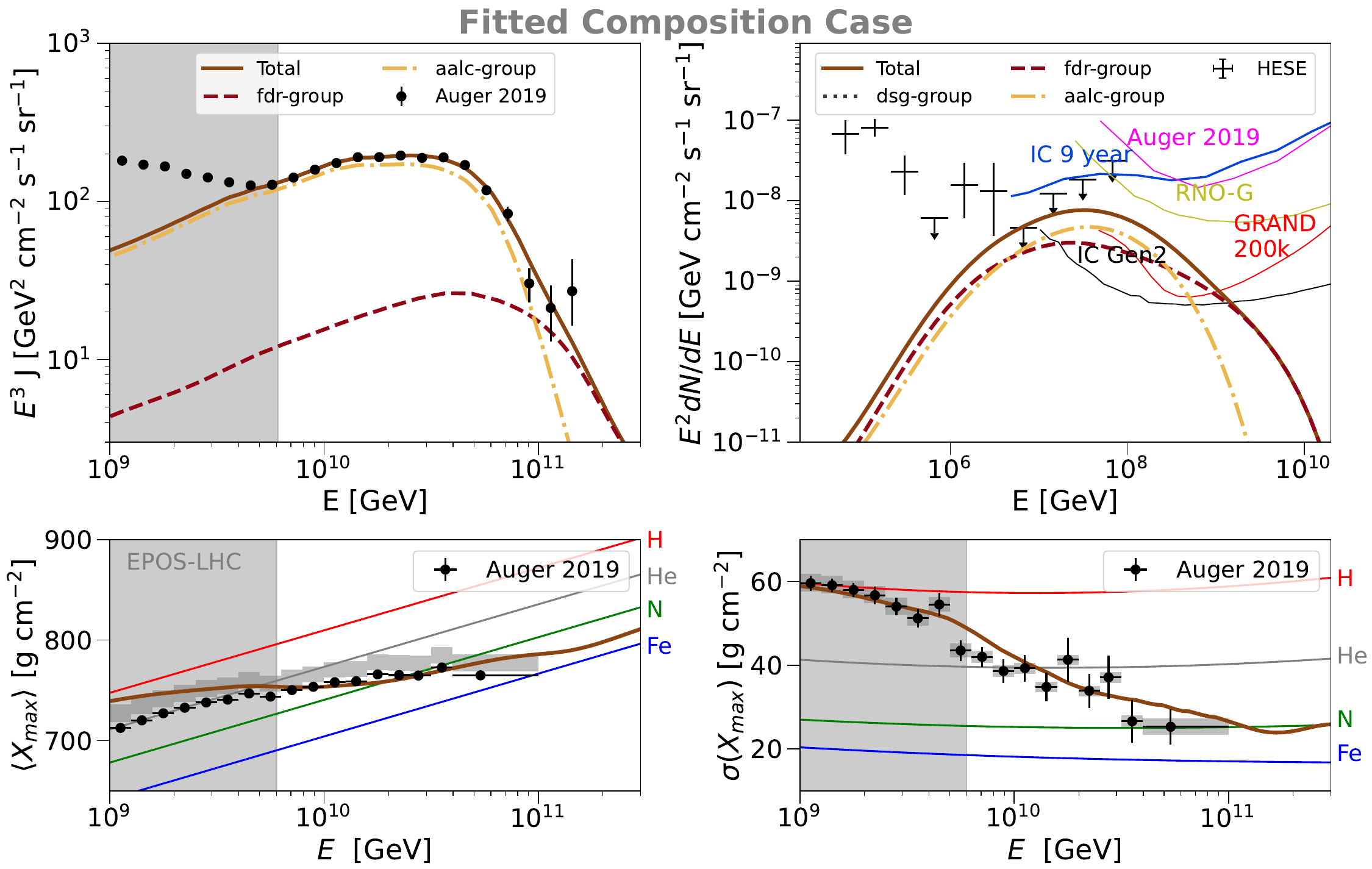} 
    \caption{Best-fit prediction for the UHECR and diffuse neutrino fluxes for the Fitted Composition Case of the Diverse Scenario, in comparison to data. Contributions from the aalc- and fdr-groups are shown with corresponding gold dashed-dotted and red dashed curves, respectively. \textbf{UHECR spectrum:} The upper left plot displays the predicted spectra against the Auger spectrum. The region where spectrum data are considered as upper limits is indicated by grey color; \textbf{UHECR composition:} The bottom plots detail the mean  ($\langle X_\mathrm{max}\rangle$, left) and standard deviation ($\sigma(X_\mathrm{max})$, right) of the UHECR composition. The grey regions mark the data range which is excluded from the fits; \textbf{neutrino diffuse flux:} The upper right plot presents the predicted (all-flavor) diffuse source neutrino flux from TDEs in comparison to the HESE neutrino data from IceCube \citep{Aartsen_2015}. In addition, we show sensitivity curves of Auger ~\citep[magenta,][]{Zas:2017Yb}, and for future neutrino telescopes,  IceCube-Gen2 ~\citep[black,][]{aartsen2019neutrinoastronomygenerationicecube}, RNO-G  ~\citep[olive,][]{Aguilar_2021}, and GRAND200k ~\citep[red,][]{lvarez_Mu_iz_2019}. \tableref{result_fitted_comp_epos} provides detailed parameter values for this best-fit case.}
    \label{fig:spectrum_epos}
\end{figure*}

In this section, we present results for the best-fit case of the UHECR-neutrino emitting TDE population, which is the Diverse Scenario in the Fitted Composition Case. In this scenario, the aalc- and fdr-groups have different parameters, such as the radius of the production region, the maximum rigidity, and the injected CR composition. \figu{spectrum_epos} illustrates the predictions for UHECR spectrum and composition and diffuse neutrino flux with a comparison to the data for the best-fit model. The obtained fitting parameters for the Diverse Scenario are detailed in \tableref{result_fitted_comp_epos}.

Our analysis shows that the predicted UHECR spectrum closely aligns with the spectrum observed by the Auger Observatory, with the fitted value of the systematic energy shift of about -9.4\%. The aalc-group primarily dominates the UHECR spectrum. Although the fdr-group's overall contribution is subdominant in spectral fitting, it plays a crucial role in composition fitting, particularly at the highest energies. The model adjustment results in a heavier composition by shifting \MeanXmax $\,$by approximately 174\%. Notably, this adjustment is consistent with recent findings from Auger, e.g., a potential systematic bias toward a heavier composition \citep{PhysRevD.109.102001}. 

In this scenario, the diffuse source neutrino flux from the TDE population does not exceed IceCube's HESE data. The contribution becomes significant at energies of $10^4$ GeV and higher, peaking at $10^8$ GeV. Unlike the UHECR spectrum, both the aalc- and fdr- groups contribute at a comparable level to the diffuse neutrino background (see the upper right panel of Fig. \ref{fig:spectrum_epos}). However, we also observe that the aalc-group produces a more peaked spectrum with a lower maximum energy in contrast to the fdr-group. As a result, the two groups predict different slopes of the neutrino spectrum in the EeV energy range. As can be seen on the upper right panel of Fig. \ref{fig:spectrum_epos}), both groups are likely detectable by forthcoming neutrino detectors like IceCube-Gen2 and GRAND200k. Regrettably, the predicted diffuse flux is too low for RNO-G to detect. 

\begin{figure*}[t]
    \centering
    \begin{tabular}{cc}
    \includegraphics[width=0.7\textwidth]{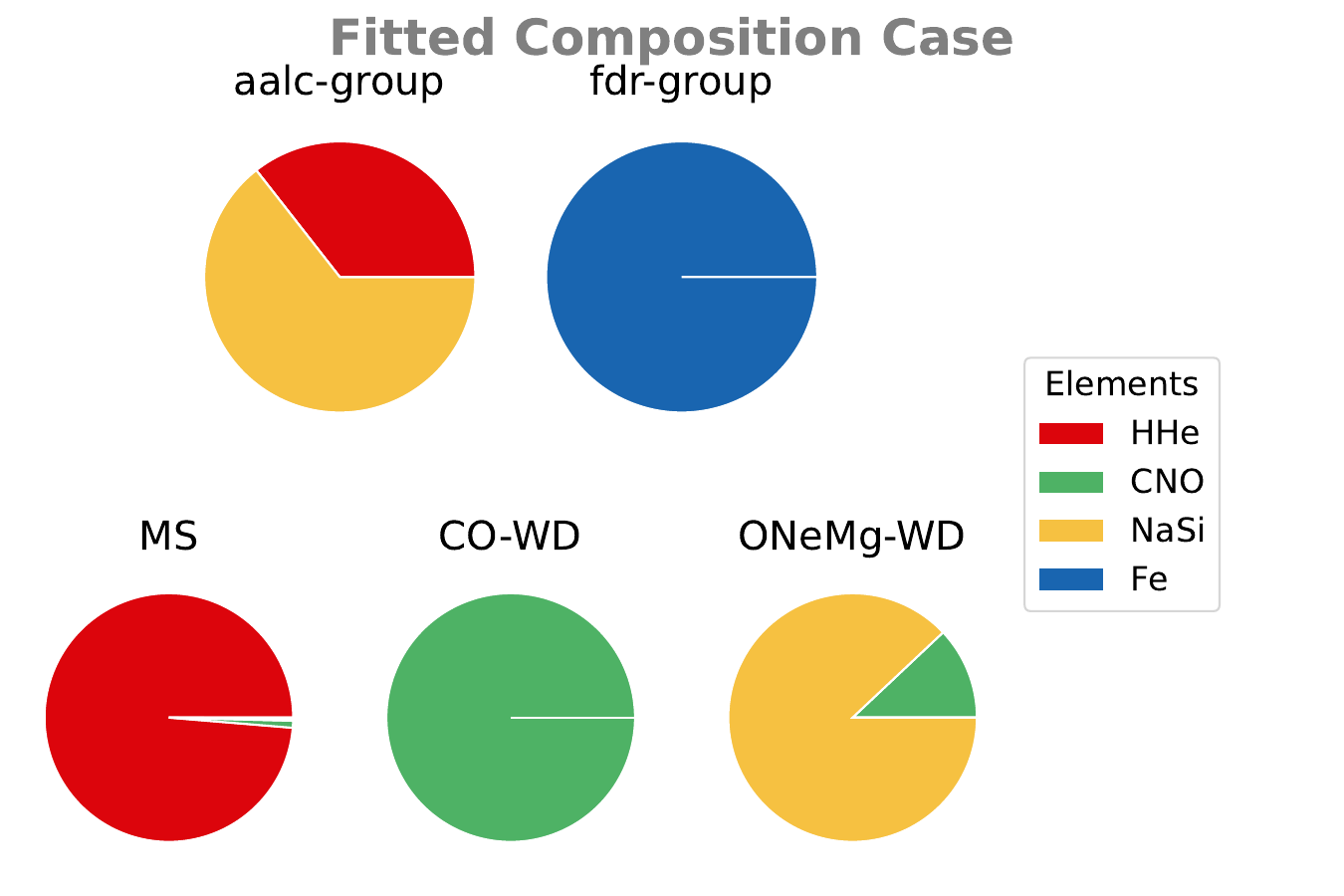} 
    \\
    \end{tabular}
    \caption{\textit{Top:} Composition of injected CRs for the best-fit case for TDEs from aalc- (left) and fdr-group (right). \textit{Bottom:} Composition of Main Sequence (MS) stars (left), CO White Dwarfs (center), and ONeMg White Dwarfs (right) for comparison. WD composition shown for comparison only; WD disruptions are unlikely for AT2019dsg, AT2019fdr, or AT2019aalc,  see discussion in \Sec~\ref{sec:discussion}.
    The elements are categorized into different groups, with each group named after the categorized elements. }
    \label{fig:mass_abundance_result}
\end{figure*}

Regarding the composition of injected UHECR, the aalc-group emits CRs in both intermediate-mass (e.g., sodium or silicon) and the light (e.g., hydrogen and helium) ranges, whereas the fdr-group mostly contributes the heavy nuclei in the best-fit scenario. This distribution is depicted in \figu{mass_abundance_result} and detailed in \tableref{result_fitted_comp_epos}. The composition fitting for the aalc-group is similar to ONeMg-WD. Still, it requires additional light elements. The absence of intermediate elements could be attributed to the loss of intermediate elements before they are transferred to the acceleration zone. Furthermore, more irons are needed for the fdr-group to make the predicted composition heavier and better fit the composition data. However, within the 1$\sigma$ AICc region, the injected composition could be pure silicon, which is similar to the ONeMg-WD.  One way to achieve an iron-abundant CR injection for both groups is by disrupting a MS star with an enhancement factor for the acceleration of heavier elements, which will be discussed in the Sec.  \ref{sec:resulst_enhancement}.

\begin{deluxetable*}{l|r|r}[htpb]
    \centering
    \caption{\textbf{Best-fit parameters for the Fitted Composition Case.} The best-fit parameters are obtained for the Diverse Scenario, considering the composition of the disrupted star as a free parameter. The analysis uses the \textsc{EPOS-LHC} air shower model. Each parameter is accompanied by its 1$\sigma$ uncertainty region, calculated for one degree of freedom (d.o.f.).}
    \label{tab:result_fitted_comp_epos}
    \tablehead{
        &\multicolumn{1}{c|}{aalc-group}
        &\multicolumn{1}{c}{fdr-group}
    }
    \startdata
     $R_{\mathrm{max}}$ [$10^9 \mathrm{GV}$]
        & $3.7_{-1.0}^{+1.4} \, 10^{9}$
        & $5.0_{-1.4}^{+1.9} \, 10^{10}$
    \\
     $r$ [$\mathrm{cm}$]
        & $1.1_{-0.3}^{+0.5} \, 10^{18}$
        & $1.1_{-0.3}^{+0.5} \, 10^{17}$ 
    \\
    $\rho_i$ [$ \mathrm{Gpc^{-3} yr^{-1} }$]
        & $194.0_{-4.4}^{+4.4}$
        & $12.9_{-2.4}^{+2.4}$
    \\
    \hline
    Injected CR fraction $f$ [\%]
        & \multicolumn{2}{c}{$\,$}
    \\
    \hline
    \multicolumn{1}{r|}{HHe}
        & $35.6_{-0.6}^{+0.6}$
        & $0.0$
    \\
    \multicolumn{1}{r|}{CNO}
        & $0.0$
        & $0.0$
    \\
    \multicolumn{1}{r|}{NaSi}
        & $64.4_{-1.0}^{+1.0}$
        & $0.0$
    \\
    \multicolumn{1}{r|}{Fe}
        & $0.0$
        & $100.0$
    \\
    \hline
     $\delta_{\mathrm{E}} (\%)$
        & \multicolumn{2}{c}{$-9.4_{-0.3}^{+0.3}$}
    \\
    $\delta_{\mathrm{\text{\MeanXmax}}} (\%)$
        & \multicolumn{2}{c}{$-172_{-18}^{+18}$}
    \\
    \hline
     $\chi^2 /\mathrm{d.o.f.}$
        & \multicolumn{2}{c}{38/25}
    \\
    \enddata
\end{deluxetable*}

In order to match the observed flux of UHECR, a local rate of $\sim$200 $\mathrm{Gpc^{-3} yr^{-1} }$ is required for the aalc-group. This value is typical for TDEs that disrupt MS stars rather than white dwarfs. For the fdr-group, the local rate should be around 10 $\mathrm{Gpc^{-3} yr^{-1} }$, which is more indicative for post-main-sequence stars.

\begin{figure*}[t]
    \centering
    \includegraphics[width=.8\textwidth]{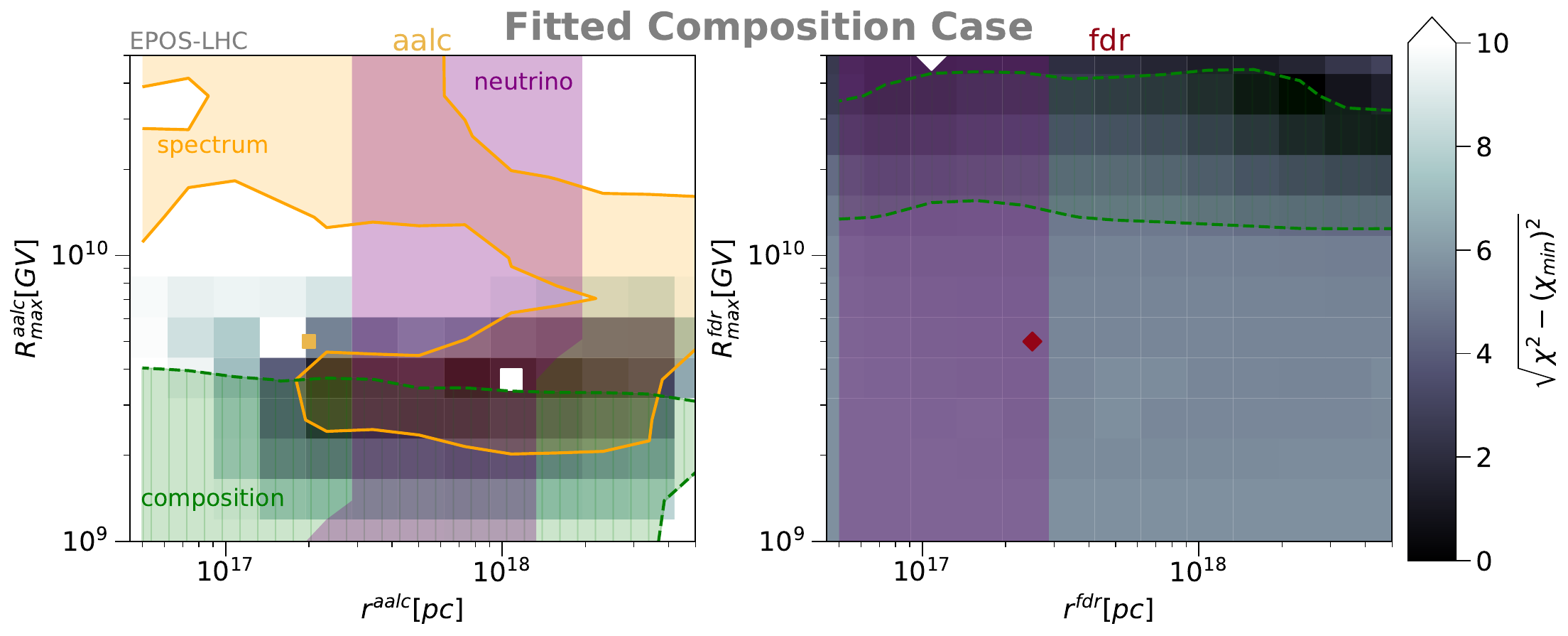} 

    \caption{The parameter space for maximum rigidity and radius of the production region for individual TDEs from the aalc- (left) and fdr-groups (right) for the Fitted Composition Case of the Diverse Scenario. The white marker indicates the best-fit parameters on each plot, as listed in \tableref{result_fitted_comp_epos}. The gold square and red diamond represent the parameters for AT02019aalc and AT2019fdr from \citet{2023ApJ...948...42W}. The yellow and green contours are the 3$\sigma$ allowed regions constrained by the spectrum and composition correspondingly. The purple contour indicates the part of the parameters space that is consistent with the expected number of neutrinos from progenitors AT2019aalc and AT2019fdr.}
    \label{fig:parameter_space_epos_fitted_comp}
\end{figure*}

\begin{figure*}[t]
    \centering
    \includegraphics[width=.65\textwidth]{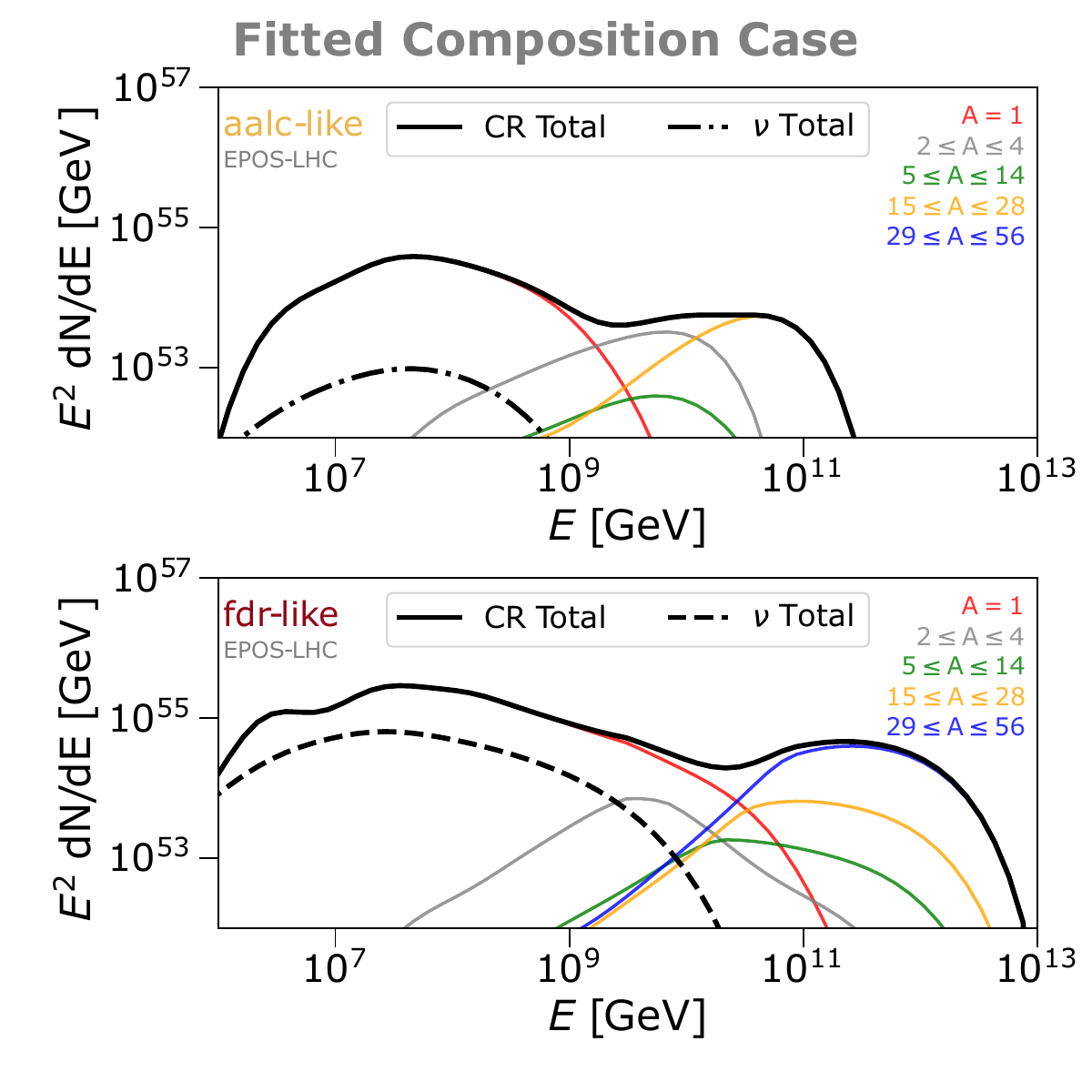} 

    \caption{The escaped spectra of CRs and neutrinos, integrated over the entire duration of TDE, from aalc- (upper) and fdr-groups (lower) for the Fitted Composition Case of the Diverse Scenario. The solid curves represent the escaped spectrum of CR, with the contribution of individual CR groups shown in different colors. The dashed-dotted and dashed curves represent the total neutrino flux.}
    \label{fig:source_spectrum_epos_fitted_comp}
\end{figure*}

Further,  \figu{parameter_space_epos_fitted_comp} explores the parameter spaces for the aalc- and fdr-groups, and  \figu{source_spectrum_epos_fitted_comp} presents the escaped spectra of CRs and neutrinos for the best-fit parameters. The aalc-group parameters are notably similar to those previously modeled for AT2019aalc \citep{2023ApJ...948...42W}, with a maximum rigidity value of $4\, 10^{9}$ GV that is expected for UHECR sources \citep{2019ApJ...873...88H, Aab:2017, Plotko_2023}. The composition 3$\sigma$ region requires values between $10^9$ and $4\, 10^9$ GV to fit the $\xmax$ data. Higher values are excluded since, in this case, the escape CRs from TDE are not dominated by a single element group but also have too many secondary light elements, contradicting the \SigmaXmax data.

On the other hand, the 3$\sigma$ allowed region, based on the spectral fitting, for the aalc-group (marked as the yellow region in the left panel of \figu{parameter_space_epos_fitted_comp}) is broad and does not significantly impact the fit. The spectrum is less sensitive to composition and does not provide a clear reason for this shape of the 3$\sigma$ spectrum allowed region. The required value of the TDE source radius of the aalc-group is $10^{18}$ cm, falling within the purple region allowed by the neutrino constraints for AT2019aalc. 

Conversely, the fdr-group requires substantially different parameters from previously used in \citet{2023ApJ...948...42W}. In order to fit the composition data, the fdr-group should have a higher maximum rigidity, with the best value of $5 \, 10^{10}$ GV. This suggests that the TDEs from the fdr-group could be a powerful UHECR source capable of accelerating iron to energies of $10^{12}$ GeV.  The preference for a high maximum rigidity arises because iron nuclei are required to contribute at energies around $5\,10^{11}$ GeV, while their presence at lower energies should be negligible. Spectral constraints are not applied to the fdr-group since its overall contribution to the UHECR spectrum is too low to significantly impact the fit.

One may wonder why in \figu{spectrum_epos} the UHECR flux for the aalc-group is significantly higher than that of the fdr-group (upper left panel), whereas the neutrino fluxes are comparable (upper right panel). As one can see in  \figu{source_spectrum_epos_fitted_comp}, fdr-like TDEs produce a (relatively) higher neutrino flux (compare dashed/dashed-dotted curves with solid black curves). The reason is the higher optical thickness due to $p\gamma$ interactions, driven by the higher luminosities (see \Tab~\ref{tab:model_parameters}) in combination with the much smaller $r$ (see \Tab~\ref{tab:result_fitted_comp_epos}).

\clearpage

\section{Enhancement of heavier nuclei in the acceleration process}
\label{sec:resulst_enhancement}

In order to further investigate the possibility of obtaining a heavier CR composition, we introduce an enhancement factor for the acceleration of heavier elements. This option is attractive in combination with the very abundant MS disruptions, which however exhibit a light progentitor composition. Realistically, nucleus selection effects may exist and cause the composition of the accelerated nuclei to differ from that of the disrupted star. One such effect might be partial disruption, which is supported by recent observations \citep{Somalwar:2023sml}. In this case, the core of the star would remain intact while the matter in the outer layers is disrupted and subsequently accreted onto the central black hole. Ultimately, the accreted material could be accelerated, resulting in the composition of the observed UHECRs. Another possibility is that there might be enhancement mechanisms in the acceleration process that would favor heavier nuclei, thus skewing the \uh\ toward heavier composition compared to the original star. 

We define an enhancement factor based on \citet{PhysRevLett.119.171101} and \citet{Hanusch_2019}, as \citet{PhysRevLett.119.171101} showed that nonrelativistic collisionless shocks can preferentially accelerate ions based on their mass/charge ratio ($A/Z^{\text{ion}}$ ratios). As a result, these shocks develop a non-thermal tail whose normalization is enhanced as $\left(A/Z^{\text{ion}} \right)^2$. \citet{Hanusch_2019} further supports the acceleration of heavier ions through hybrid simulations, which reveal that the injection efficiency of ions into the diffusive shock acceleration process depends strongly on the shock Mach number. Their simulations showed that the injection efficiency for ions with high $A/Z$ ratios is enhanced. 
To generalize the idea, we assume that injection efficiency is proportional to $\left(A/Z^{\text{ion}} \right)^\alpha$, where $\alpha$ is the enhancement factor. Additionally, we assume that thermal particles lose an electron ($Z^{\text{ion}}=1$) before being transferred to an acceleration zone to maximize the effect. The net effect is that the abundance of the element $i$ in the accelerated stream is scaled by a power of the mass number, leading to new, rescaled mass fractions, $f_i^{\mathrm{new}}$: 
\begin{align}
   f_i^{\mathrm{new}} = \frac{(A_i)^{\alpha}} X_{i}{\sum_{i=1}^N (A_i)^{\alpha}} X_{i}. 
\end{align}

\begin{figure*}[t]
    \centering
    \begin{tabular}{c}
    \includegraphics[width=0.85\textwidth]{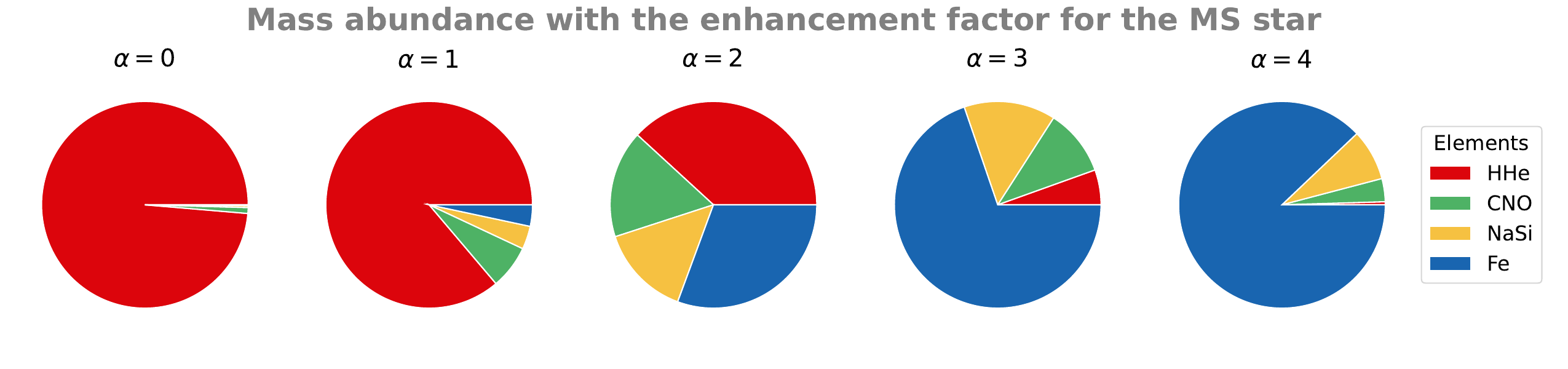}
    \end{tabular}
    \caption{The mass abundance profiles across different enhancement factor values for the MS star. The elements are categorized into different groups, with each group named after the categorized elements.}
    \label{fig:mass_abundance_evolved}
\end{figure*}

To ensure that the total mass available for each element (``nuclear budget'') is not exceeded, we calculate the required mass of the injected elements and compare it to the initial mass available in the disrupted star. We will mention if constraints from the nuclear budget require lowering the UHECR luminosity from individual TDE. In \figu{mass_abundance_evolved}, we show the rescaled fractions for our representative stellar compositions, depending on $\alpha$, ranging from $\alpha=0$ (no scaling) to $\alpha=4$. 

\begin{figure*}[htpb!]
    \centering
    \begin{tabular}{c}
    \includegraphics[width=.7\textwidth]{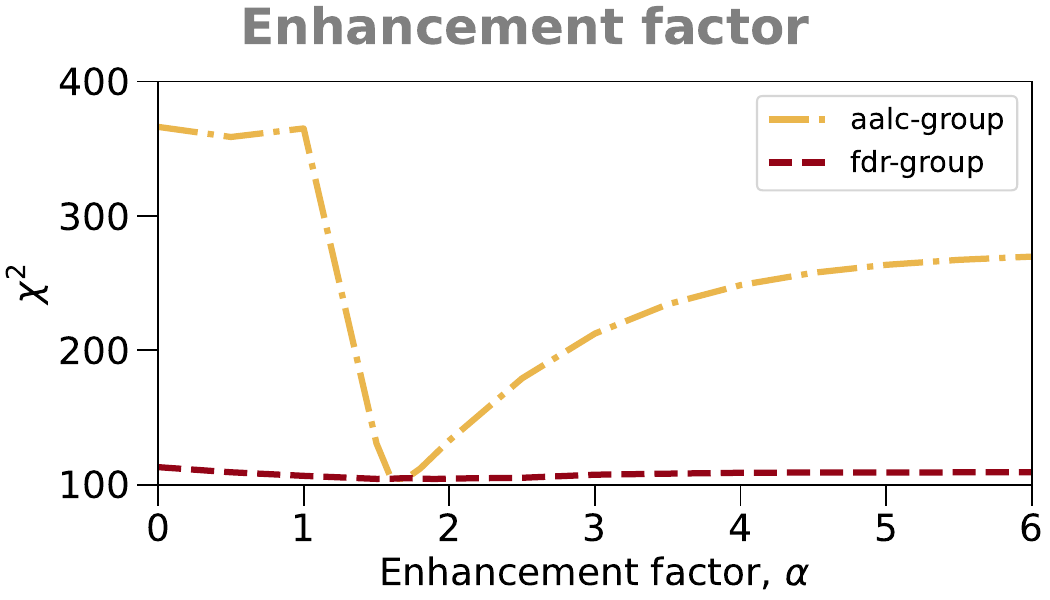}
    \end{tabular}
    \caption{ $\chi^2$ values as a function of the enhancement factor $\alpha$ of aalc- and fdr- group. For each enhancement factor value of a group, a dedicated fit is performed to determine the minimum $\chi^2$ value while minimizing all other fitted parameters, including the enhancement factor of the other group. }
    \label{fig:enh_chi2_value}
\end{figure*}

In our analysis, we use the Diverse Scenario and focus on the fixed composition of a MS star with various values of $\alpha$ for injected CRs. In this scenario, we assume that the aalc- and fdr-groups have different $\alpha$ values. For each combination of $\alpha_{\mathrm{aalc}}$ and $\alpha_{\mathrm{fdr}}$, we conduct a fit to determine the minimum $\chi^2$ value while minimizing over all other fitted parameters. We show $\chi^2$ as a function of the enhancement factor $\alpha$ in \figu{enh_chi2_value}.  For the aalc-group, the fit requires the value of $\alpha$ to be between 1.5 and 2. For the fdr-group,  there is a slight preference for values of $\alpha = 1.9$  for the fdr-group, but not statistically significant due to the small contribution of the fdr-group to the UHECR spectrum. 

Overall, the Enhancement Factor Case is excluded by more than 8$\sigma$ compared to the Fitted Composition Case due to the nature of the enhancement factor and the assumed fixed composition. The model predicts a light composition for values of the enhancement factor near zero, which contradicts the UHECR composition observations. As the value of $\alpha$ increases, the model becomes too efficient at accelerating heavy elements. However, to explain UHECR data, an intermediate-mass group is needed. Different assumptions could help address this issue. For example, there may be two groups of TDEs: one that disrupts white dwarfs with intermediate mass elements and another that disrupts MS stars with an enhancement factor for acceleration of heavier elements. This approach would provide both lighter and heavier elements, potentially improving the fit of the composition data. However, such a model would introduce too many free parameters and could be statistically excluded in favor of the simpler Fitted Composition Case.

\begin{figure*}[t]
    \centering
    \begin{tabular}{c}
    \includegraphics[width=.7\textwidth]{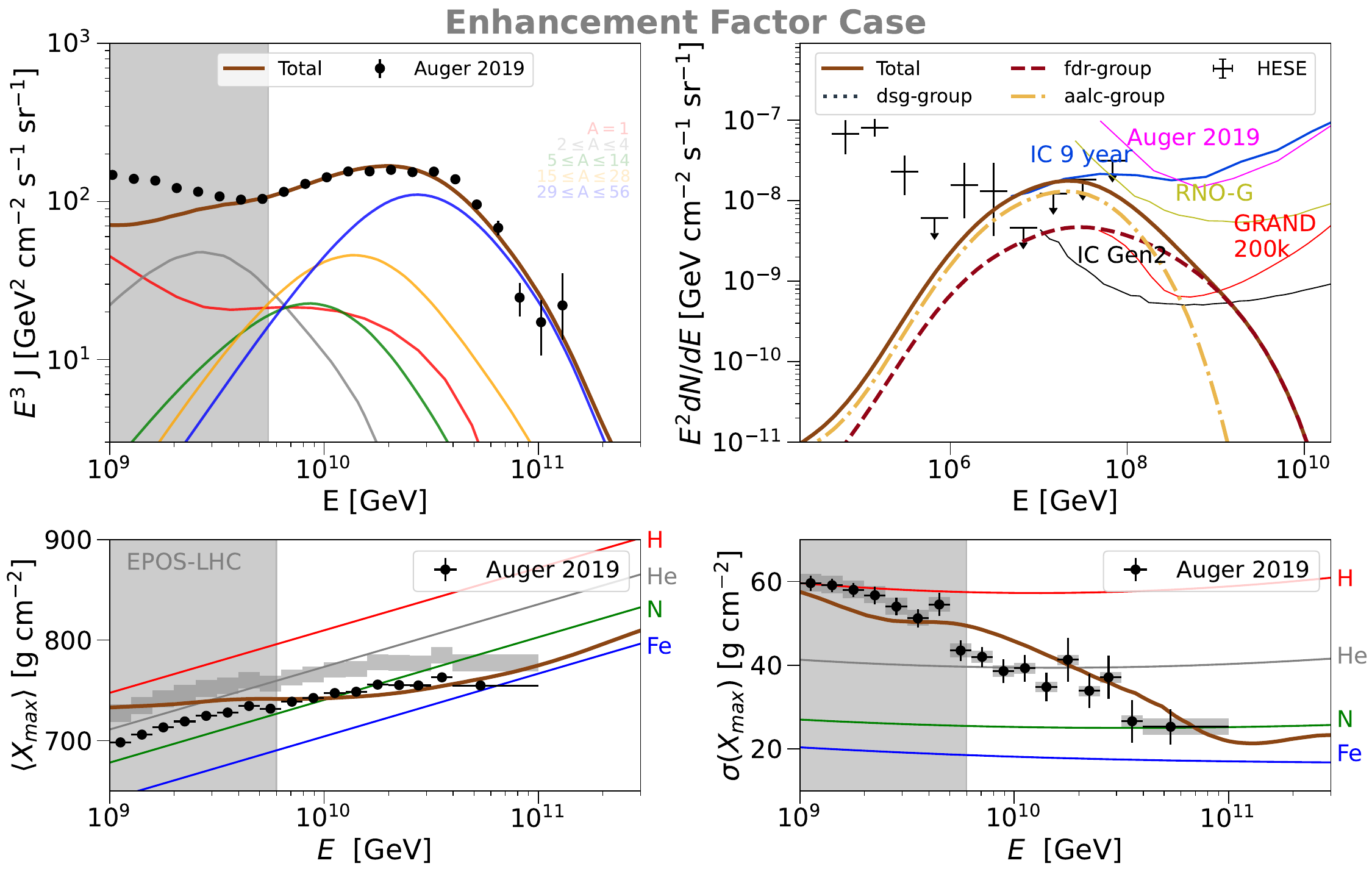}
    \end{tabular}
    \caption{The best-fit prediction for UHECR and diffuse neutrino flux for the Enhancement Factor Case in comparison with the data. See caption of  \figu{spectrum_epos} for details. \tableref{result_fitted_comp_epos} provides detailed parameter values for this best-fit case.}
    \label{fig:enh_factor_spectrum}
\end{figure*}

\figu{enh_factor_spectrum} compares the best-fit case to the Auger spectrum and composition, along with the predicted diffuse neutrino flux, for the Enhancement Factor Case. The corresponding parameters of the best-fit case are presented in \tableref{result_enh_epos}. While the model's predictions can explain the UHECR spectrum, the main contribution comes from the iron group, which contradicts the current UHECR composition observation. Additionally, the diffuse neutrino flux from the aalc-group exceeds the HESE limits to a high local rate of 1116 $ \mathrm{Gpc^{-3} yr^{-1} }$. These two factors result in a poor quality of fit. One potential solution to improve the fit could be using a more complex enhancement factor, such as having iron elements lose more electrons than the silicon group, resulting in a more silicon-dominant composition.

\begin{deluxetable*}{l|r|r}[htpb!]
    \centering
    \caption{\textbf{The best-fit parameters for the Enhancement Factor Case.} The best-fit parameters obtained from analyzing the Diverse Scenario of TDE populations assume the disruption of the Main Sequence (MS) star with an enhanced acceleration factor for heavier elements. The analysis uses the \textsc{EPOS-LHC} air shower model. Due to the poor fit, we present only the best value of the parameters without uncertainty.}
    \label{tab:result_enh_epos}
    \tablehead{
        &\multicolumn{1}{c|}{aalc-group}
        &\multicolumn{1}{c}{fdr-group}
    }
    \startdata
     $R_{\mathrm{max}}$ [$10^9 \mathrm{GV}$]
        & $1.0\, 10^{9}$
        & $2.6\, 10^{10}$
    \\
     $r$ [$\mathrm{cm}$]
        & $7.3\, 10^{17}$
        & $1.6\, 10^{17}$ 
    \\
    $\rho_i$ [$ \mathrm{Gpc^{-3} yr^{-1} }$]
        & $1116$
        & $3$
    \\
    $\alpha$ 
        & 1.5
        & 1.9
    \\
    \hline
    Injected CR fraction $f^{\mathrm{new}}$ [\%]
        & \multicolumn{2}{c}{$\,$}
    \\
    \hline
    \multicolumn{1}{r|}{HHe}
        & $55.2$
        & $43.8$
    \\
    \multicolumn{1}{r|}{CNO}
        & $15.1$
        & $16.5$
    \\
    \multicolumn{1}{r|}{NaSi}
        & $11.2$
        & $13.4$
    \\
    \multicolumn{1}{r|}{Fe}
        & $18.6$
        & $26.3$
    \\
    \hline
     $\delta_{\mathrm{E}} (\%)$
        & \multicolumn{2}{c}{$-8.3$}
    \\
    $\delta_{\mathrm{\text{\MeanXmax}}} (\%)$
        & \multicolumn{2}{c}{$-300$}
    \\
    \hline
     $\chi^2 /\mathrm{d.o.f.}$
        & \multicolumn{2}{c}{104/24}
    \\
    $\Delta \mathrm{AICc} $
        & \multicolumn{2}{c}{$> 8\, \sigma$}
    \\
    \enddata
\end{deluxetable*}

\clearpage

\section{Discussion}\label{sec:discussion}

Let us now take a broader view of our results and discuss their significance, implications, and limitations. 
\\

\underline{\emph{Choice of parameters, degeneracies.}} 
In this study, we fixed several key parameters, such as the SMBH mass, the dissipation efficiency, the minimum energy, and the spectral index of injected nuclei. We also took into account that certain parameters -- including the SMBH mass and the  IR, OUV, and X-ray luminosities --  are constrained by observations.   For instance, we fixed the non-thermal cosmic ray luminosity to be \(L_A = \varepsilon_{\mathrm{diss}} \, \dot{M} \, c^2\), where the mass accretion rate \(\dot{M}\) is taken to be proportional to the Eddington luminosity (and therefore to the SMBH mass). We then derived the local TDE rates that are required to power the observed UHECR flux.  These parameters are, however, somewhat degenerate. Indeed, lower $\varepsilon_{\mathrm{diss}}$ can be compensated by a higher local rate or a higher \(L_A\)  (higher SMBH mass).
Similarly, a higher minimum nuclei energy or a harder injection spectrum lowers the energetic requirement as more non-thermal power is allocated at higher energies, see discussion in \citet{2024arXiv240219314W}. Due to this degeneracy, our derived local rates should only be interpreted as indicative. Moreover, degeneracies should be taken into account when comparing with other works; for example, 
\citet{2018NatSR...810828B} only fit the product of local rate and baryonic loading.
\\

\underline{\emph{Cosmological TDE rates, connection with jetted TDEs.}} 
Despite the existing degeneracies, 
it is interesting to compare the TDE rates found in this work with those expected from theory and observations. 
The theoretically expected  local rate of all TDEs,
$\rho_{\mathrm{0}} \sim (0.7 - 20) \, 10^{3}~\mathrm{ Gpc^{-3} yr^{-1}}$ (depending on the minimal SMBH mass; see \Sec~\ref{sub:tderates}) serves as an upper limit for the rate of the sub-class of UHECR-emitting TDEs.  Our fit results are consistent with such limit: 
we found $\rho_{\mathrm{0}} \sim 200$ yr\(^{-1}\)Gpc\(^{-3}\) for the aalc-group and $\rho_{\mathrm{0}} \sim10$ yr\(^{-1}\)Gpc\(^{-3}\) for the fdr-group, which amount to a fraction $\eta \sim 5 \, 10^{-4} - 0.3$ of the maximum value.

It is noteworthy that our TDE fraction range is roughly compatible with the fraction of jetted TDEs, $\eta \sim 10^{-4}$ to $10^{-2}$~\citep{2023JCAP...11..049P}, which suggests a possible connection with the existence of jets in TDEs.  In the model of \citet{2023ApJ...948...42W}, the jet, which can be completely off-axis, may be the accelerator;  the non-thermal UHECRs isotropize in the larger region enclosed by the dust torus, thus making the source effectively a quasi-isotropic emitter. On the other hand, a jet directly pointing towards us, such as in the model of \citet{2021NatAs...5..472W}, is expected to lead to significant electromagnetic signatures (e.g., non-thermal X-rays) which have not been observed in the three cases of neutrino-coincident TDEs but may account for the luminous X-ray afterglows of jetted TDEs, such as the recently reported AT 2022cmc \citep[e.g.,][]{2022Natur.612..430A,2023NatAs...7...88P,2024arXiv240611513Y,2024arXiv241107925Y}.  
Earlier approaches, such as \citet{2018NatSR...810828B}, also used jetted TDEs for the combined description of UHECR and neutrino data. Note that from the UHECR perspective, there is no conceptual difference between those earlier models and our approach, as UHECRs are deflected by extragalactic magnetic fields and, therefore, do not directly trace the original direction of the jet. In fact, the value of the local (apparent) rate of jetted TDEs that was assumed in \citet{2018NatSR...810828B},   0.1 yr\(^{-1}\)Gpc\(^{-3}\), corresponds - after taking into account the beaming factor (opening angle $\theta_j \simeq 0.1$ from \citet{2023JCAP...11..049P}) -- to a local TDE rate $\rho_{\mathrm{0}} \sim$ 200 yr\(^{-1}\)Gpc\(^{-3}\), which agrees very well with the values we found for the aalc-group. 

For neutrinos, however, a jetted model may lead to a higher flux at Earth, due to the contribution of neutrinos from internal shocks inside the jet, which can be very efficient. This additional flux comes from jets pointing in our direction, such as in  \citet{2018NatSR...810828B}.

\underline{\emph{Type of disrupted stars, nuclear composition.}} 
We found that the best-fit composition for the aalc-group is a mix of medium- and light-mass elements. It is overall similar to that of ONeMg-WD (in accordance with \citet{Zhang:2017hom}), but requires additional light elements. The latter feature might be due to the loss of intermediate elements before they are transferred to the acceleration zone; it could also indicate the existence of two subgroups within the aalc-group, one involving white dwarfs and the other main-sequence stars. However, such a model is beyond the scope of our current analysis with the available data. For the fdr-group, the composition is likely dominated by iron or silicon, within the 1$\sigma$ AICc uncertainty region. Such a composition is also broadly consistent with an ONeMg-WDs. 

While the composition analysis favors attributing the bulk of the UHECR flux to the disruption of WDs,  several other facts weigh in favor of the disruption of MS stars instead. They are:  
\begin{itemize}
    \item The local TDE rates. Our required rates for the aalc-group are well consistent with MS stars, but somewhat in tension with the predicted rates of disruption of WD stars (unless the most extreme values of TDE-WD rates are invoked, see Sec. \ref{sub:tderates}). 

    \item The estimated black hole masses. In the current understanding of TDEs white dwarfs can only be disrupted by intermediate-mass black holes with typical masses around \(10^5\) $\msun$, while all three observed TDEs involve supermassive black holes around \(10^7\) $\msun$. Based on that, Disruption of the WD is unlikely for AT2019dsg, AT2019fdr, and AT2019aalc. Therefore, for these TDEs, the MS star is a more probable candidate.
    

    \item The estimated masses of the disrupted stars for AT2019aalc and AT2019fdr, which are of several $\msun$, in the range of MS stars, see Table \ref{tab:model_parameters} and \citet{2023ApJ...948...42W}.  
\end{itemize}

 There are potential solutions to this tension, where additional physical effects bring the required nuclear composition into consistency with the disruption of MS stars.  A possibility is a strong enhancement of heavier elements through the injection processes. A simple, power-law enhancement factor might reproduce the required iron-like composition for the fdr-group 
 (see Sec. \ref{sec:resulst_enhancement}), whereas for the aalc-group, a more complicated process that favors medium-mass elements would be required.  
  Another idea is that during the disruption of a main-sequence star, central densities, and temperatures can dramatically increase for a very short period, leading to significant nuclear burning. However, there is no clear consensus on whether these conditions are sufficient:  \citet{1989A&A...209...85L} argue in favor, while \citet{2009ApJ...705..844G} come to a negative answer.
Thus, a self-consistent picture has not emerged yet. Such a picture will require a more comprehensive (UHECR-emitting) population model, which would not rely as much on the observed neutrino-TDE coincidences. Indeed, these may not be representative of the whole population; rather, they could be outliers with extraordinarily large neutrino fluences. 

During the completion of this work, a new study appeared \citep{2024arXiv240817419M},  revealing a re-brightening of AT2019aalc four years after its discovery. This new flare has characteristics -- such as a dust echo and X-ray emission -- that are similar to the first one. While the classification of AT2019aalc as  TDE still remains unclear, it is speculated that its peculiar properties might be explained by the phenomenology of TDEs in AGN, for example, partial disruption. Spectroscopically, the flare source can be classified as Bowen Fluorescence Flare  \citep[BFF, a relatively new class of flaring AGN,][]{Makrygianni_2023}. Interestingly, another such BFF with a very similar emission spectrum, AT2021loi, has been identified and has been found to be within the (rectangular) 90\% confidence region of neutrino event IC-230511. This neutrino arrived 680 (290) days after the first (second) optical peak of AT2021loi.\footnote{In fact, an additional neutrino-flare pair was found, involving  AT2017bgt and the neutrino IC-200410A, which was, however, a poorly reconstructed cascade event.} These observations indicate that, indeed, AT2019aalc might be exceptional and that our aalc-group may correspond to a special type of TDEs in AGN related to BFFs, including AT2019aalc, AT2021loi, and, possibly, AT2017bgt. The spectroscopic results indicate that intermediate to heavy elements, required to describe the UHECR data, are indeed available in the system, even though a direct link to our TDE model requires more information on the origin of the emission lines, which is not yet available. Nevertheless, the limited sample size, classification ambiguities, and potential AGN contamination warrant caution against drawing strong conclusions from these individual events.

\underline{\emph{Compatibility between observed neutrino energies and predicted neutrino spectra.}}
The direct connection between neutrino energy and CR rigidity $E_\nu \simeq 0.025 \, R \, e$ (cf., discussion after \equ{rmax}, applied to nuclei heavier than hydrogen) means that neutrinos are produced at energies following $R_{\mathrm{max}}$ --  much higher than the inferred reconstructed neutrino energies in the 100~TeV range for the three neutrino-TDE associations.
A quantification of the statistical compatibility between our predicted neutrino spectra (see \eg\ \figu{spectrum_epos}, upper right panel) and the energy uncertainty of the IceCube events is nevertheless not possible because uncertainties for the three TDE-associated events have not been published so far, and
(especially for muon tracks) the energy estimate is strongly event-dependent and non-Gaussian; it depends \eg\ on the uncertainty on the vertex distance  and how much energy was lost before reaching the detector (which means it depends on the direction, depth, etc.). 

An additional challenge is that the detector response depends on the spectral shape itself. This can be seen, for instance, in the neutrino association of IC-170922A with the flaring AGN blazar TXS 0506+056 for a similar event, where the most probable neutrino energy was given to be 290~TeV: the harder the assumed spectrum in that energy range, the more the probability distribution prefers higher reconstructed neutrino energies \citep[see Suppl. Materials, Fig. S2]{2013Sci...342E...1I}; our predicted spectrum is even much harder in that energy range. A generic attempt to quantify these uncertainties for power law spectra has been made in \citet{2023ApJ...948...42W} (see e.g. gray-shaded areas in Fig.~10).

This problem has been, in fact, studied in greater detail for AGN blazars in \citet{2024A&A...689A.147R}, where the predicted neutrino energy follows the maximal proton energy -- which is frequently used as free parameter in AGN blazar models: if it is chosen to be in the few-PeV range, it can produce neutrino spectra peaking in the 100~TeV energy range. At these energies, the target photon number is however low, which means that super-Eddington proton luminosities need to be invoked to produce reasonable neutrino event rates, which are incompatible with the standard accretion paradigm in AGN. A higher proton energy can significantly reduce this tension, which leads to the hypothesis that AGN blazar neutrino spectra may also peak at higher energies. \citet{2024A&A...689A.147R} have therefore analyzed the detector response for these spectral cases in greater detail (see App.~A) and have come to the come to the conclusion that for $p\gamma$-like spectra much higher neutrino energies are compatible with the detector response (see Fig. A.2). In  fact, the predicted spectra in \citet{2024A&A...689A.147R} therefore peak at around the same energies as our TDE predicted spectra.  While the theoretical arguments are different here, the conclusions on the detector response apply similarly. 
However, in \citet{2021JCAP...10..082O} the probability that a 20~PeV neutrino is reconstructed with a muon energy as low as 210~TeV was estimated for the individual event IC-190730A associated with an AGN blazar to be around 25\%. While such an individual event is consistent with the hypothesis that the spectrum peaks at higher energies, it raises the question if three independent events from three different astrophysical objects with such a spectrum can be found at such low energies, which may be at the level of $0.25^3 \simeq 2\%$ following their arguments. This line of argumentation would, however, also apply to set of AGN blazars, which raises two more generally applicable possibilities: 1) The observed neutrinos (with the exception of KM3-230213A perhaps, see below) are not coming from a population of sources producing the UHECRs or 2) Some systematics on the detector side leads to an offset in the reconstructed neutrino energies.

Another hint for hard TDE spectra comes from stacking searches, where too soft spectra limit the possibility of neutrino-TDE associations \citep{2023arXiv230715531N}.

\underline{\emph{Broader connections: individual extreme UHECR and neutrino events.}} 
Regarding the derived UHECR energies, the fdr-group might be able to produce particles with high energies, such as the \(2.2 \, 10^{20}\) eV particle observed in 1991 \citep{Bird_1995} and the Amaterasu particle with an energy of \(2.4 \, 10^{20}\) eV in 2021 \citep{doi:10.1126/science.abo5095}. We note that \citet{Unger_2024} suggests that a transient source may produce the Amaterasu particle, which positions TDEs as a potential candidate. Moreover, \citet{bourriche2024localvoidcomprehensiveview} proposed that the Amaterasu particle could have originated from emitted iron nuclei, which aligns with our results for the composition of the fdr-group. Recently, the detection of a high-energy neutrino event KM3-230213A by KM3NeT-ARCA, with a median neutrino energy of 220~PeV (range 72~PeV to 2.6~EeV at the 90\% CL)  was reported \citep{KM3NeT:2025npi}, with no clear source association. This energy range is compatible with our diffuse flux computation in \figu{spectrum_epos}, even though our spectrum peaks at slightly lower energies. It is interesting that the interpretation of this event coming from a transient with a lifetime comparable or shorter than the live time of the experiment (335 days) alleviates the tension between IceCube and KM3NeT data \citep{Li:2025tqf,Neronov:2025jfj} -- compared to other possible explanations, such as cosmogenic neutrinos.

\underline{\emph{Fit quality and possible developments.}} 
The best $\chi^2 /\mathrm{d.o.f.}$ value obtained in our analysis was approximately 2.5, which indicates that the fit quality is not ideal. This suggests that while our model captures the broad characteristics of the data, there are likely residuals and complexities in the data that are not fully accounted for. However, attempts to further improve the fit by introducing additional parameters or making the model more complex are unlikely to yield significant benefits. The reason is that the current data may not provide enough additional information to justify an increased model complexity, and such adjustments might only make the theoretical model overly intricate and challenging to interpret. Even in more straightforward cases, such as those discussed by \citet{Heinze:2019} and \citet{Aab:2017}, achieving a perfect fit has proven challenging.

A potential development of our analysis might be including information on the arrival direction of the observed UHECRs. Considering that this exercise has turned out to be difficult even with simple source models, as indicated by  \citet{Capel_2019} and \citet{Abdul_Halim_2024}, we feel that applying it to our case of a realistic source model might introduce excessive complication while not adding significant improvements. 

Taking a broad look at the situation, we feel that substantial advancements will be possible in the future, when more extensive and detailed multimessenger observations of TDEs become available. These will make it possible to produce more realistic models of TDEs as as a population -- or mix of populations -- of UHECR and neutrino sources. Those models can then be combined with similarly detailed and robust methods of statistical analysis, to produce state-of-the-art fits of all the available data.   

\section{Summary and conclusions}\label{sec:conclusions}

For several years, TDEs have been discussed as potential emitters of UHECR and high energy neutrinos. 
A significant advancement occurred recently, with the proposed associations between neutrinos detected at IceCube and the TDEs/candidates, AT2019dsg, AT2019fdr, and AT2019aalc.  All of these events have two common features.  The first commonality are  strong dust echoes, due to the absorption of optical-UV and X-ray photons by dust and consequent re-emission in the IR band. The peak of the IR flare is observed as delayed relative to the blackbody peak, by the travel time from the central source to the dust region, $\sim {\mathcal O}(10^2)$ days. Remarkably, using IR echoes as a selection criterion has led to the identification of AT2019aalc and its associated neutrino. The second commonality is that the coincident neutrinos have all been detected with  similar delays of hundreds of days, which place them close to the dust echo peaks. These two facts suggest that the dust echoes may contribute significantly to neutrino production, by providing the target photons for hadron-photon scattering where neutrinos originate. Intriguingly, the primary nuclei must have energies in the UHECR range to exceed the photo-pion production threshold off the infrared targets, therefore the neutrino-TDE associations might be interpreted as indirect evidence for the acceleration of UHECRs in TDEs. In light of these associations, we have revisited the idea that TDEs may power the observed diffuse UHECR flux. 

We extended a quasi-isotropic neutrino production model, which was previously proposed to interpret the neutrino-TDE associations, by injecting a variety of nuclear species and allowing the relevant parameters (size of production region and maximal rigidity) to vary. We combined that with a UHECR transport code to predict the cosmic ray composition at Earth. Finally, we performed a fit to  the UHECR spectrum and composition data from Auger, and, for each set of fitted parameters, we computed the predicted diffuse neutrino flux at IceCube. It is important to note that the expected local rate of TDEs is determined by fitting the normalization (required emissivity) of the computed cosmic ray spectra to UHECR data, and that predicted diffuse neutrino fluxes are an output (rather than an input) of our calculation. 

 The main challenges for this work have been: (i) extrapolating from a few neutrino-emitting TDEs to a population of UHECR-emitting TDEs, (ii) reconciling the favored local TDE rate with the nuclear injection composition required by the data, and (iii) ensuring consistency with the individual neutrino-TDE associations. 
 We have identified a  best-fit scenario where two populations of TDEs, one being AT2019aalc-like and the other AT2019fdr-like, that describe all relevant UHECR data best while meeting existing constraints. 
In such a scenario, the aalc-group has an injection composition that consists of a mix of light and intermediate (Na-Si) isotopes, and dominates the UHECR spectrum. In contrast, the  fdr-group is characterized by  a Fe-like composition and reproduces the highest energy tail of the spectrum. Therefore, our results indicate that at least two different classes of TDEs contribute to the observed UHECR flux. The inferred parameters (radius of production region and maximal rigidity) are found to be in consistency with earlier works. The expected local TDE rates range between $\sim$10 and $\sim$100 yr\(^{-1}\)Gpc\(^{-3}\), where a significant uncertainty is allowed by degeneracies in the parameters. These rates correspond to a fraction of the overall TDE rate, which, interestingly, is compatible with the fraction of jetted TDEs.

 From our results, it is not clear what type of stars must be disrupted to power the observed UHECRs. White dwarf disruptions seem to have the right mass composition,  but are too rare and associated with lighter black hole masses. Main sequence stars are preferred as progenitors because their local TDE rate is high enough, and the estimated black hole masses and star's masses of AT2019fdr, and AT2019aalc favor this hypothesis. However, their composition (in terms of the mass fraction) is light. 
As an option to resolve this discrepancy,   we have proposed an enhancement of the injection of heavier elements in the acceleration process, which means that these elements should exhibit larger non-thermal contributions compared to the lighter nuclei. We found that the simplest realization of this idea could make the data compatible with main sequence star disruptions, but a more complicated enhancement mechanism that is most effective for intermediate-mass elements (rather than heavy elements) would be needed to provide a good fit. During completion of this work, evidence that AT2019aalc belongs to a special type of events (which can be identified with Bowen Fluorescence Flares -- BFF) was presented,  which might be related to (partial) TDEs in AGN. New candidate neutrino associations involving members of this group were identified. It is remarkable that our UHECR data analysis leads to the same conclusion, that the aalc-group might be a special class, thus substantiating the idea of two distinct stellar populations or classes of events further.

In order to associate a specific astrophysical class with UHECRs, reproducing the observed spectrum and composition at Earth is a necessary condition. However, direct source identification remains challenging, as UHECRs are not only significantly deflected by intergalactic magnetic fields but also experience long propagation time delays \citep{mbarek2025propagationdelaysultrahighenergycosmic}. These delays make it impossible to directly associate UHECRs with transient sources like TDEs. Instead, source and propagation simulations provide a valuable approach to investigating whether certain source classes, such as TDEs, can reproduce the observed UHECR spectrum and composition. At the same time, these simulations help identify key challenges and constraints on the physical conditions required for TDEs to be viable UHECR sources.

We note that the predicted neutrino flux associated with UHECRs inevitably peaks at $\sim$10-100~PeV, whereas the observed neutrino energies had much lower inferred energies in the 100~TeV range; see \Sec~\ref{sec:discussion} for a discussion of the compatibility.
 
On the other hand, such high peak energies suggests that a TDE or a TDE-like AGN flare \citep{Yuan:2025zwe} could be the source of the recently observed ultra-high-energy neutrino by KM3NeT-ARCA \citep{KM3NeT:2025npi}. Future radio detection instruments like IceCube-Gen2 ~\citep[black,][]{aartsen2019neutrinoastronomygenerationicecube}, RNO-G  ~\citep[olive,][]{Aguilar_2021}, and GRAND200k ~\citep[red,][]{lvarez_Mu_iz_2019}, should be able to robustly test our prediction for the neutrino flux.

In conclusion, we find it overall plausible that TDEs could power the UHECRs in the light of recent neutrino-TDE associations. However, the extrapolation from a few neutrino-emitting TDEs, which may not be archetypic and may be also not very statistically confident in terms of their directionality, to a population of UHECR-emitting TDEs is non-trivial due to limited knowledge and possible selection effects -- and some of the parameters, such as the observed neutrino energies, do not support this hypothesis. Independent of that, TDEs are interesting UHECR candidates for different reasons, such as their transient nature, high enough emissivity, and negative source evolution, and our UHECR model and analysis (including the diffuse neutrino flux prediction) remain valid independent of the neutrino associations. Future data will potentially improve the predictions, both on the TDE side and the neutrino side.

\begin{acknowledgments}
We would like to thank Nadine Bourriche, Francesca Capel, Ralph Engel and Marek Kowalski for useful discussions.  PP was supported by the International Helmholtz-Weizmann Research School for Multimessenger Astronomy, largely funded through the Initiative and Networking Fund of the Helmholtz Association. C.L. aknowledges funding from the NSF, through the grants 2012195 and 2309973.  

This work made use of the following software packages: \texttt{Jupyter} \citep{2007CSE.....9c..21P, kluyver2016jupyter}, \texttt{matplotlib} \citep{Hunter:2007}, \texttt{numpy} \citep{numpy}, \texttt{python} \citep{python}, \texttt{scipy} \citep{2020SciPy-NMeth, scipy_13352243}, and \texttt{iminuit} \citep{iminuit, James:1975dr, iminuit_13902219}. Software citation information aggregated using \texttt{\href{https://www.tomwagg.com/software-citation-station/}{The Software Citation Station}} \citep{software-citation-station-paper, software-citation-station-zenodo}.

The authors acknowledge the use of \texttt{Grammarly}\citep{grammarly2024} and \texttt{OpenAI's ChatGPT}  \citep{openai2024chatgpt}  for language refinement and proofreading. 

\end{acknowledgments}

\bibliography{ref}{}
\bibliographystyle{aasjournalv7}

\end{CJK*}
\end{document}